\documentclass[11pt]{article}
\usepackage{graphicx,color}
\usepackage{caption}
\usepackage{setspace}
\begin{document}
\title{Possible new phase transition in the 3D Ising Model associated with boundary percolation}
\author{Michael Grady
\\Department of Physics\\ State University of New York at Fredonia\\
Fredonia NY 14063 USA\\
grady@fredonia.edu}
\date{\today}
\maketitle
\thispagestyle{empty}
\begin{abstract}
In the ordered phase of the 3D Ising model, minority spin clusters are surrounded by a boundary of dual plaquettes.  As the temperature is raised, these spin clusters become more numerous, and it is found that eventually their boundaries undergo a percolation transition when about 13\% of spins are minority.  Boundary percolation differs from the more commonly studied site and link percolation, although it is related to an unusual type of site percolation that includes next to nearest neighbor relationships. Because the Ising model can be reformulated in terms of the domain boundaries alone, there is reason to believe boundary percolation should be relevant here.  A symmetry-breaking order parameter is found in the dual theory, the 3D gauge Ising model.  It is seen to undergo a phase transition at a coupling close to that predicted by duality from the boundary percolation.  This transition lies in the disordered phase of the gauge theory and has the nature of a spin-glass transition. Its critical exponent $\nu \sim 1.3$ is seen to match the finite-size shift exponent of the percolation transition further cementing their connection.   This predicts a very weak specific heat singularity with exponent $\alpha \sim -1.9$.  The third energy cumulant fits well to the expected non-infinite critical behavior  in a manner consistent with both the predicted exponent and critical point, indicating a true thermal phase transition. Unlike random boundary percolation, the Ising boundary percolation has two different $\nu$ exponents, one associated with largest-cluster scaling and the other with finite-size transition-point shift.  This suggests there are two different correlation lengths present.
\end{abstract}
\linespread{1.2}
\newpage
\section{Introduction}
The Ising models in two and three dimensions are the most basic spin models which undergo order-disorder transitions.  These have been extremely well studied and, of course, an exact solution exists in the two-dimensional case. The 3D model has always been a bit more of a mystery, and in this paper we explore the possibility of a weak secondary phase transition within the ordered phase.  Presumably this is associated with some geometrical change in spin-clustering, but the exact nature of this reordering is unknown.  The situation is clearer in the dual theory, the 3D gauge Ising model.  Here the suspected transition is in the disordered phase, and clearly has the nature of a spin-glass transition. In other words we have identified a symmetry-breaking order parameter in the dual theory, but not in the Ising model itself.  However,  the Ising model does exhibit an interesting percolation phenomenon near the critical point predicted by duality from the spin-glass transition.   This is a percolation of the domain boundaries between + and - spin clusters.  As shown below, boundary percolation can be considered a third type of percolation, beyond site and link percolation, although it has a close relationship to an unusual type of site percolation.  Of course percolation is not always related to a phase transition, but sometimes it is.  It's linkage in this case to a symmetry-breaking transition in the dual theory provides strong evidence that it is associated with a phase transition in this case.  The argument is further strengthened by independent fits of the third energy cumulant to consistent critical behavior about the suspected critical point.  Each investigation, the order-parameter in the dual theory, the boundary percolation finite-size shift, and the third energy moment, yields an independent determination of the critical exponent $\nu$, all of which agree to a fairly close tolerance.  
  
 In the following, first the boundary percolation concept is fleshed out and studied in both the 2D and 3D Ising models, as well as for 3D random percolation.  It is found that the latter has the same critical exponents as for site percolation, and in fact is equivalent to site percolation if next to nearest neighbors are included in the cluster definition.  The 3D Ising case is particularly interesting in that an analysis of the finite-size shift in percolation threshold gives a critical exponent very different from the percolation value, even though the cluster scaling still obeys the percolation exponents.  This suggests it is linked to a phase transition with its own dynamical scaling and correlation length.  Then we move on to the dual theory and introduce the spin-glass order parameter.  A Monte Carlo study here shows clear crossings in the Binder cumulant and second-moment correlation length divided by lattice size.  Correlation-length finite-size scaling is exhibited around the suspected critical point using scaling collapse plots, which also yield critical exponents.  The critical exponent $\nu$ is found to match well with that found from the finite-size shift in percolation threshold.  Finally, we study energy moments, such as specific heat and higher moments.  Unlike an order parameter, these have both critical and non-critical pieces, so fitting can be difficult. This leads to the selection of the third energy cumulant as the best prospect for finding critical behavior as it can be fit without a non-critical part other than a constant.  A Monte Carlo study with several times $10^{9}$ sweeps per point on  $30^3$, $40^3$, and $50^3$ lattices yields a precise determination of this quantity. An independent critical behavior fit in the region of the suspected critical point gives values for both $\nu$ and $\kappa _c$ which agree well with the two other predictions.  A substantial jump in the coefficient of the critical scaling fit across the transition further cements evidence for a thermal singularity here. 

The rather high value of $\nu \sim 1.3$ gives a highly negative value for the specific heat exponent $\alpha = 2-d\nu  \sim -1.9$. This means that both the specific heat and third cumulant have finite singularities.  A very weak infinite singularity is expected in the fourth cumulant and stronger ones in fifth and higher.  In the Ehrenfest classification this transition is fourth order.  We attempted to measure fourth and fifth cumulants to find evidence of peaks growing with lattice size as expected from infinite singularities, but even with the rather large sample size here, these were still largely obscured by random error.  However, finite singularities are just as singular as infinite ones, so perhaps one lesson is that one should not necessarily obsess over trying to find infinite singularities in transitions of such high order.
\begin{figure}[th!]\centering
                    \includegraphics[width=0.4\textwidth,  clip]{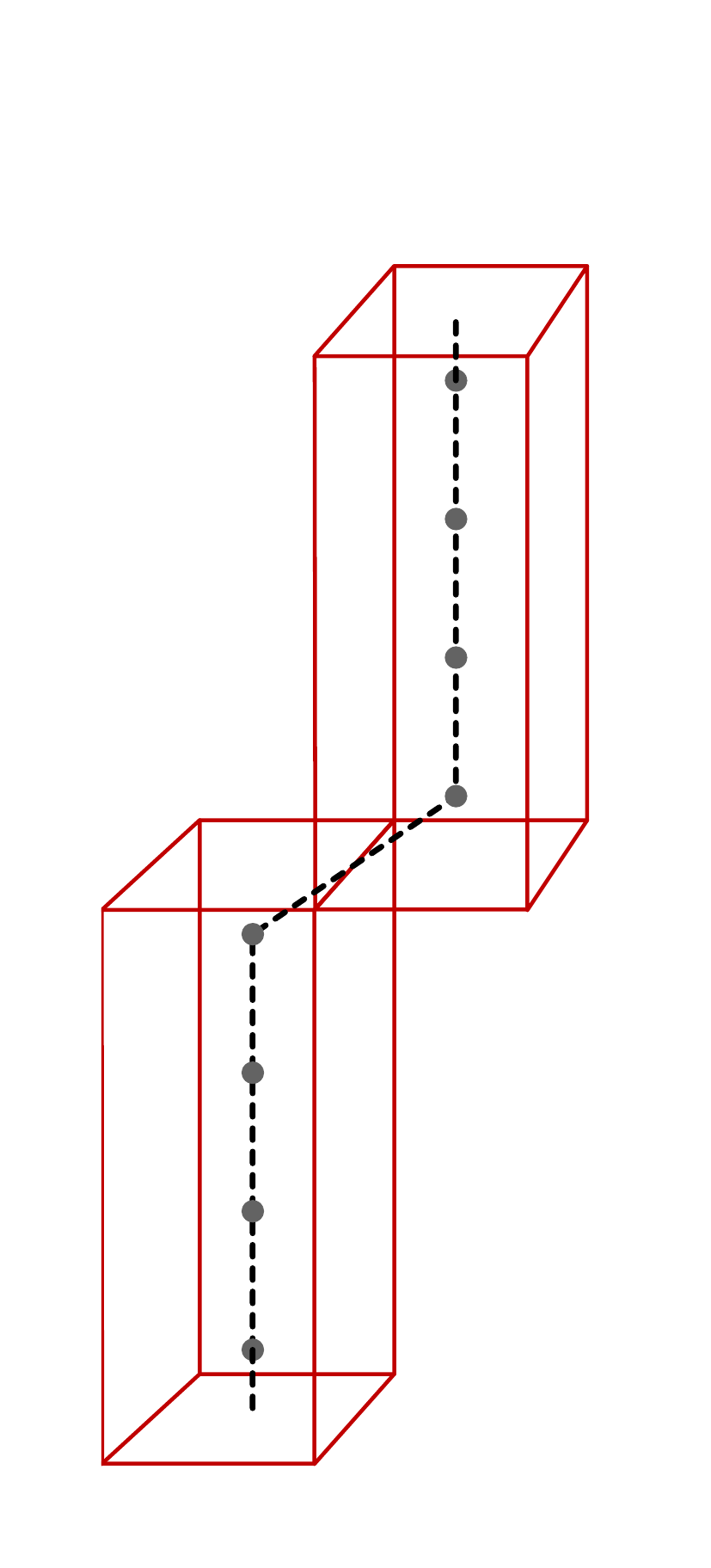}
                                                  \caption{Example of a boundary cluster. Sites marked with dots have opposite orientation to all surrounding sites.}
          \label{fig1}
       \end{figure}

\section{Boundary percolation}
The standard partition function for the Ising model is
\begin{equation}
Z=\sum _{\{\sigma \}}\exp (\kappa \sum _{n.n.}\sigma _i \sigma _j) ,
\end{equation}
where the $\sigma $'s are classical spins taking values $\pm 1$ and the coupling is between nearest neighbors only.
There is a well-known reformulation of the Ising models in terms of the boundaries themselves\cite{feynman}. This reformulation even leads to an alternate exact solution in the 2D case\cite{kac}. The partition function can be written
\begin{equation}
Z=\sum _A N(A)\exp (-\kappa A)
\end{equation}
where $A$ is the total area of dual boundary plaquettes (or dual boundary links in 2D) in a configuration and N(A) are the number of distinct non-intersecting boundary configurations with that area. In this formulation there are no spins or domains. Only the boundary surfaces need exist, and the entropy associated with these surfaces controls the phase transition.  This is one reason why percolation of the domain boundary  might be important for this model, as opposed to, say, site percolation. For instance, the density of states could change abruptly 
when an infinite boundary cluster forms, because for a finite cluster the area
is usually an increasing function of the volume, whereas an infinite cluster can easily
grow in volume without adding much to the area. If this were the case then the free energy
would form a singularity at the boundary percolation point. 

We define boundary percolation as follows.
Consider the set of all boundary links that connect + and - sites. These are each 
associated with a plaquette on the dual lattice. These plaquettes form closed surfaces
separating clusters of + and - spins. These surfaces can form clusters themselves, if
we define boundary clusters to be made up of boundary surfaces that share dual-lattice
links. For instance, Fig.~1 shows a single boundary cluster. This same configuration would,
however, count as two separate site-clusters, since sites are clustered only along lattice
directions. If the site cluster concept is extended to include sites connected by 
face diagonals, i.e next nearest neighbors (NNN) in addition to nearest neighbors (NN),
then these redefined site clusters
would appear to coincide with the boundary cluster concept.  Indeed we have verified for thousands of configurations
that those with percolating boundaries also have percolating NN+NNN site-clusters,
and vice versa, so they do appear to measure the same thing.

 It seems boundary percolation, which in three dimensions could also be called plaquette percolation,  
has only been studied before in the form of the 
equivalent extended NN+NNN site percolation problem\cite{nn+nnn}, as a part of surveys
of various extended percolation models, but never applied to the Ising model. The
main result of these studies is establishing a threshold for random NN+NNN percolation at
a minority site probability of 0.1372(1). Because the Ising model is interacting, correlations
would be expected to modify this result, but still it should be kept in mind.
  Fig.~2ab shows the evolution of the Ising model boundary percolation threshold $\kappa _{L}^{*}$
with lattice size $L$ for both two and three dimensions. These both scale well with
the finite-size scaling relation 
\begin{equation}
\kappa_{L}^{*}=\kappa _c - cL^{-1/\nu }
\end{equation}
where $\kappa _c $ is the infinite lattice threshold. The percolation threshold is
defined here as the point where 50\% of lattices have a cluster which percolates in 
all directions. For two dimensions, boundary percolation exists in the random phase, and
ceases in the ferromagnetic phase. The above fit gives $\kappa _c =0.4405(5)$ which agrees well with the known ferromagnetic
transition point $\frac{1}{2}\ln (\sqrt{2}+1)\simeq 0.44069$. This is just the opposite of majority-site percolation, which
happens only in the magnetized phase. Thus in two dimensions boundary-percolation 
and site-percolation seem equally relevant. The exponent derived from the finite-size scaling fit to Fig.~2a is 
$\nu = 1.261(18) $. This seems slightly different from the standard 2D site-percolation 
exponent
$\nu =4/3$ but of course that is for a non-interacting system.  There could also be a small correction from next to leading order scaling effects. As far as we know,
the critical exponents for random NN+NNN site percolation or random boundary-percolation have not been previously measured. In principle they could differ
from site percolation, however in three dimensions we find below that the critical exponents for random boundary percolation appear to be the same as for ordinary site percolation.  Probably the same is true in two dimensions, but we did not perform that measurement.
\begin{figure}[th!]\centering
                    \includegraphics[width=0.49\textwidth,  clip]{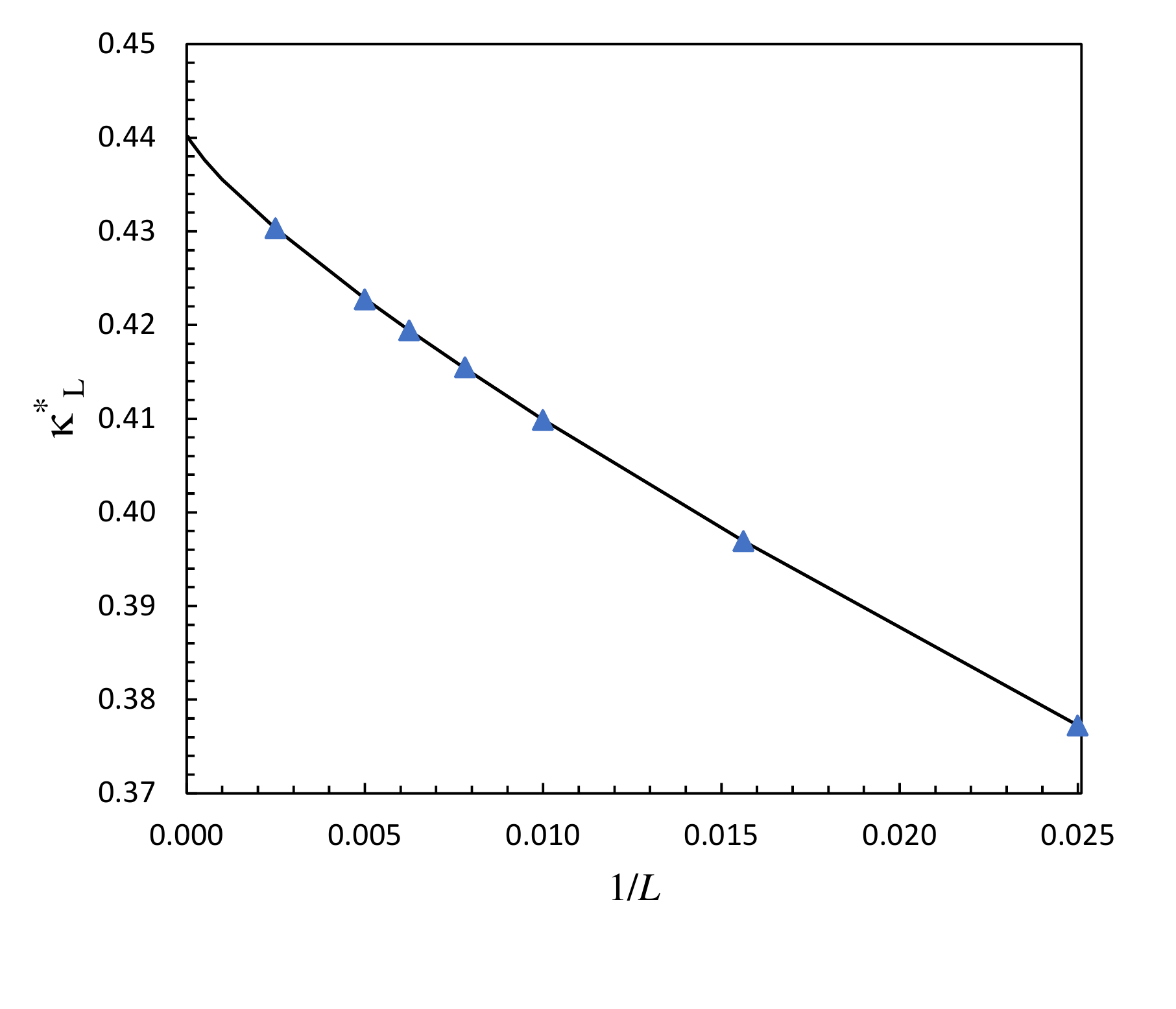}
                    \includegraphics[width=0.49\textwidth,  clip]{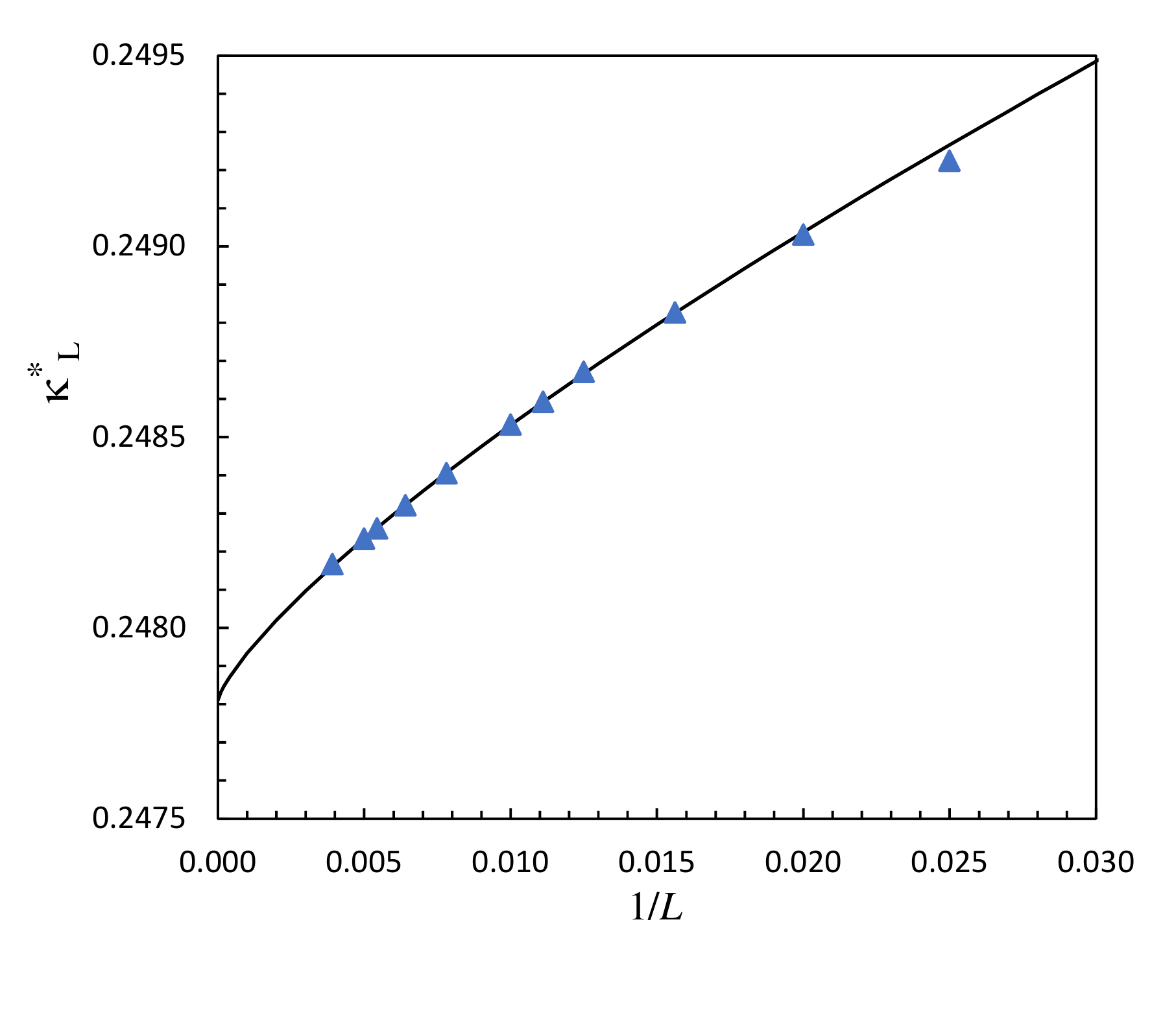}
                                                  \caption{Finite-size shift of the boundary percolation threshold for the 2D (a) and 3D (b) Ising model. Error bar ranges are 1/10 to 1/20 of symbol size.}
          \label{fig2}
       \end{figure}

In two dimensions minority sites never
percolate. In three dimensions there are a lot more paths. Majority sites always 
percolate and minority sites percolate in the random phase and about the first 5\%
of the ferromagnetic phase, measured by temperature. Eventually minority sites get too few and percolation is 
lost at $\kappa = 0.2346(13)$ where the magnetization is 
about 0.62\cite{mk}. There is no visible effect on
other quantities at this point. For comparison, the ferromagnetic transition is at $\kappa = 0.2216595(26)$ \cite{fl}.  Because in boundary percolation the clusters are more liberally defined, 
it persists even further into    
the ferromagnetic phase. From the fit to Fig.~2b we find $\kappa _c = 0.24781(4)$ and 
$\nu = 1.30(3)$. Here the magnetization is about $0.7364$. This means that 
13.18(4)\% of the sites are minority, which is about 4\% lower than the value mentioned above, $p_c =0.1372$,
for random NN+NNN site percolation (our study of random boundary percolation  below also corroborates this value).
The value found here for the exponent $\nu $ is particularly interesting.  It is not at all close to the correlation length exponent for random site percolation $\nu \simeq 0.88$\cite{wang,staufferbook} measured from the finite-size scaling of the infinite cluster.  For random boundary percolation we find a similar value below. Even in the 3D Ising case where interactions could change the result we still find scaling of the largest cluster gives $\nu \simeq 0.87$ (detailed below).  We are led to conclude that a different correlation length is controlling the finite-lattice shift exponent in this case.  This makes the case of boundary percolation in the 3D Ising model of considerable theoretical interest, because a system with two different correlation lengths diverging at the same place is, to say the least, unusual.

\begin{figure}[th!]\centering
                    \includegraphics[width=0.5\textwidth,  clip]{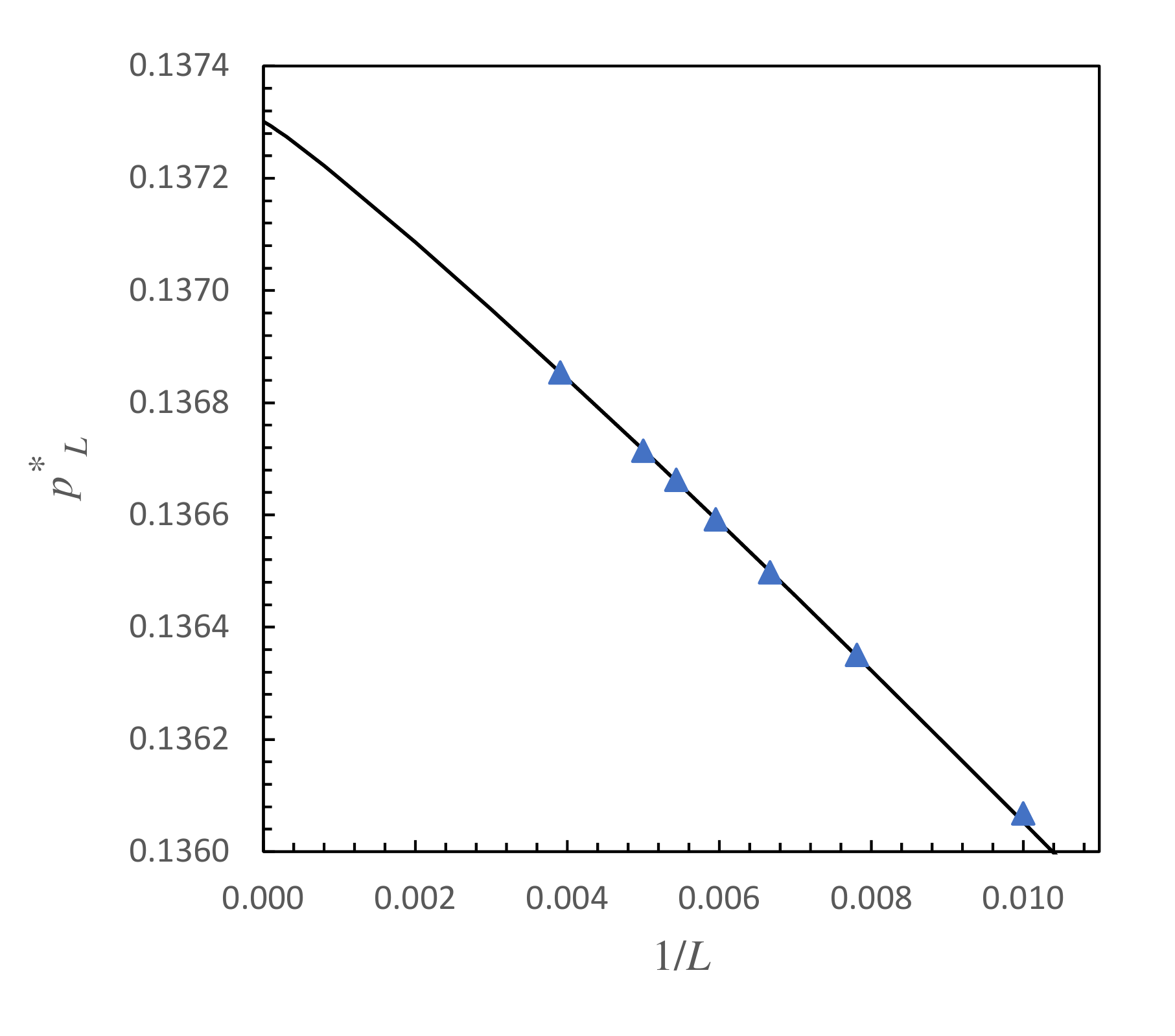}
                                                  \caption{Finite-size shift of boundary percolation threshold for 3D random percolation model. Error bar range is about 1/8 symbol size.}
          \label{fig3}
       \end{figure}

Now we consider the case of random boundary percolation.  This is an interaction-free model where positive sites are placed at random in the lattice, with the fraction of positive sites given as $p$. The remaining sites are, of course, set negative.  Fig.~3 shows the finite-size shift of percolation threshold, $p_L ^*$.  From the scaling relation given above we find the infinite lattice threshold as $p_c =0.13730(4)$, which is fairly close to the percentage of positive sites at the 3D Ising boundary percolation threshold (they differ by 4\%). However, the exponent here is quite different from the 3D Ising value of $1.30(3)$. We find $\nu=0.91(5)$, consistent with well-known measurements of the  site-percolation exponent.  One can also determine $\nu$ from scaling of the largest cluster.  If one defines $P$ to be the fraction of plaquettes occupied by the largest cluster, then the same finite-size scaling analysis as is usually applied to the magnetization in a magnetic system undergoing a thermal phase transition can be applied \cite{binderbook} to $P$, its susceptibility 
\begin{figure}[bh!]\centering
                    \includegraphics[width=0.49\textwidth,  clip]{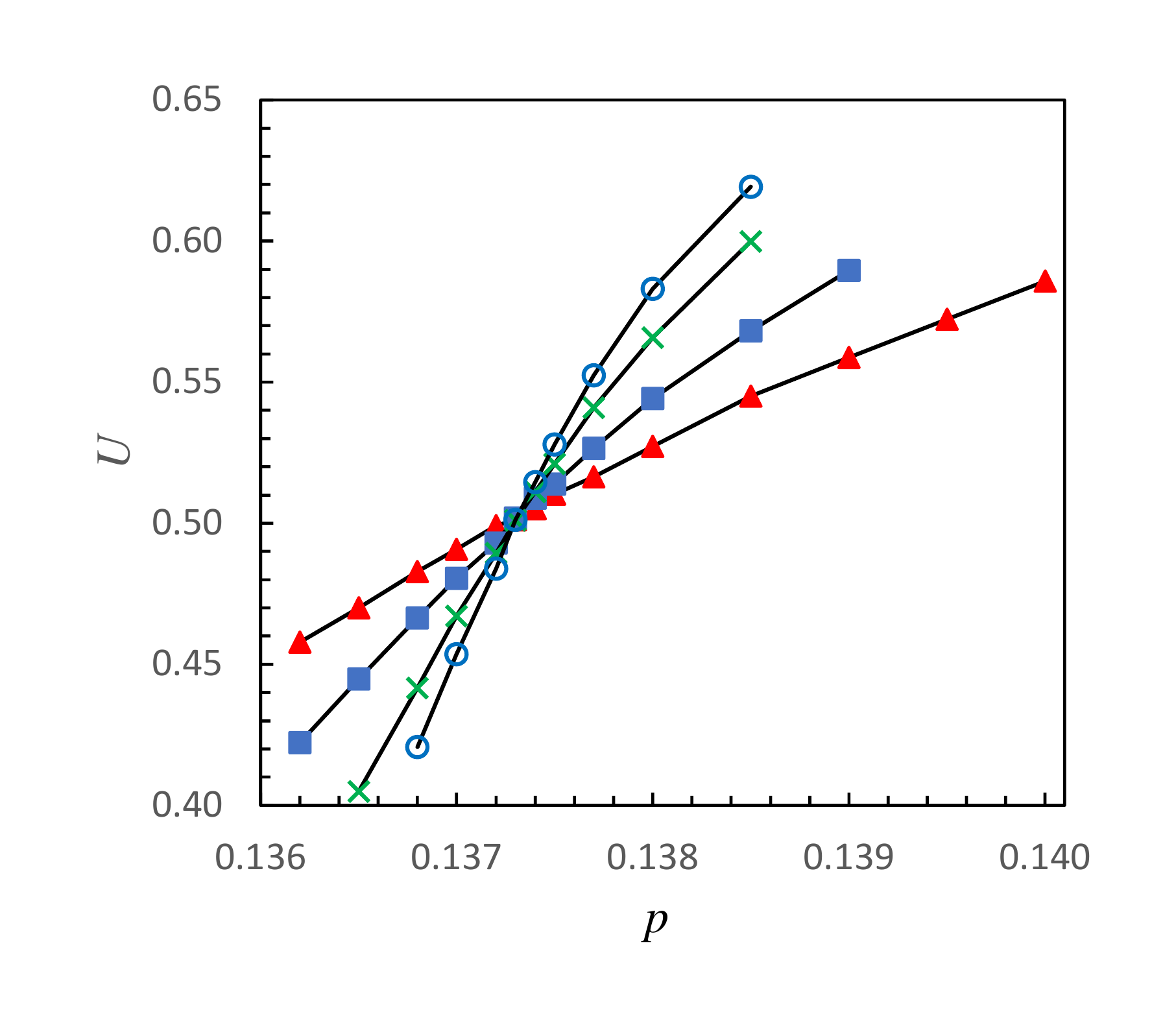}
                         \includegraphics[width=0.49\textwidth,  clip]{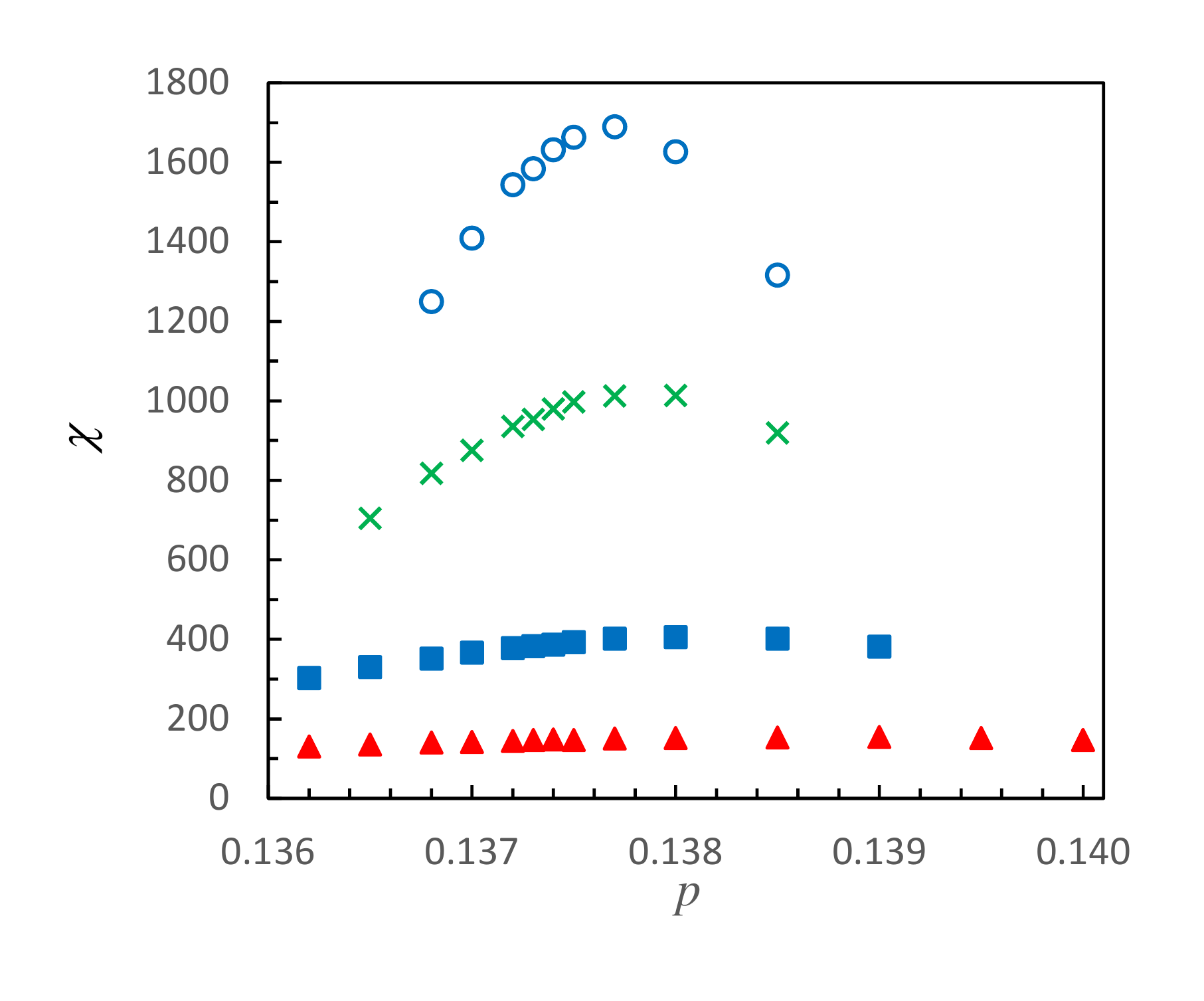}
                         \includegraphics[width=0.49\textwidth,  clip]{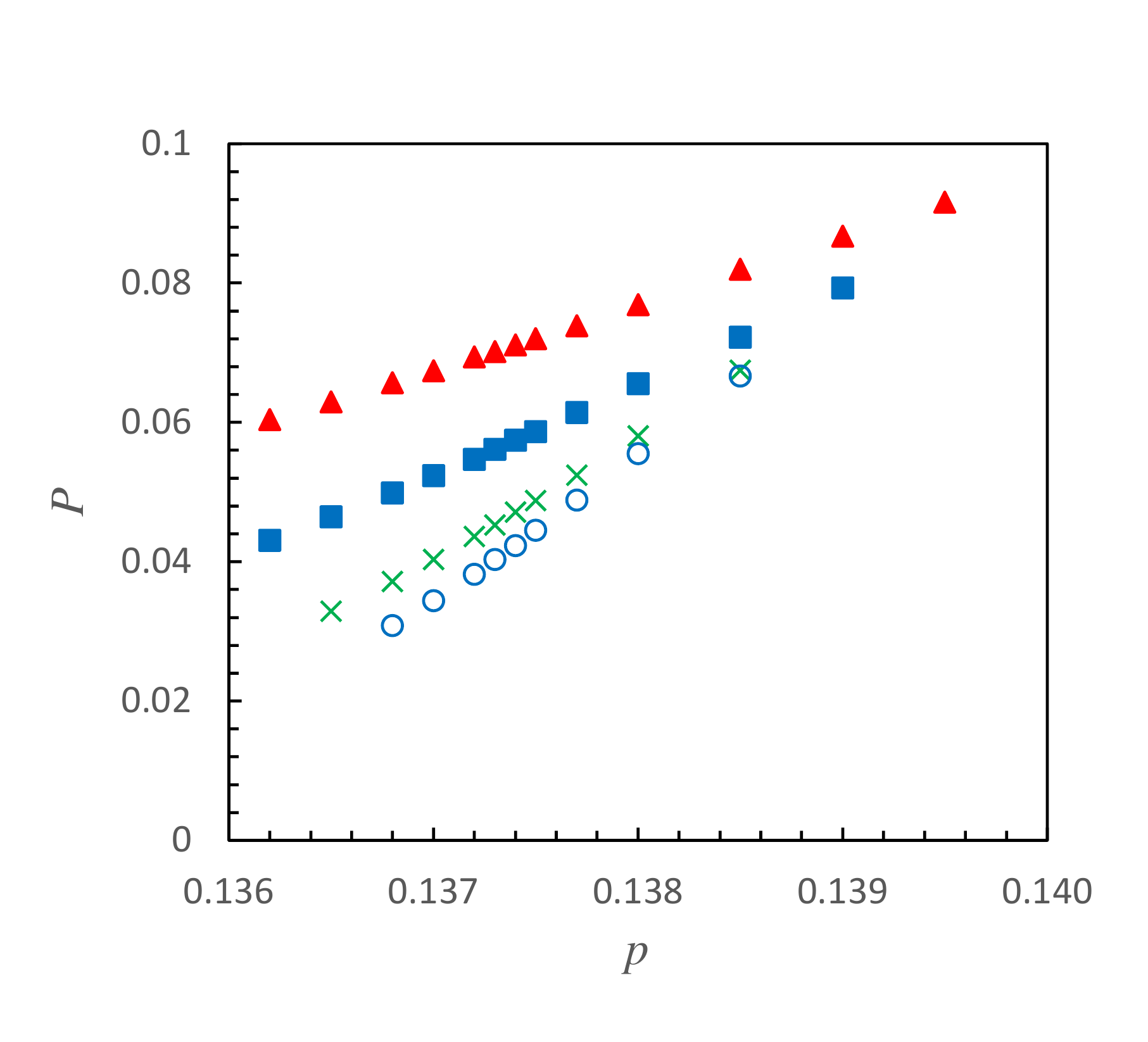}
                                  \caption{Boundary percolation study in the 3D random percolation model. Binder cumulant $U$ (a), susceptibility $\chi$ (b), and fraction of sites occupied by largest cluster $P$ (c) vs. fraction of positive sites, $p$. Triangles are $40^3$, boxes $64^3$, $\times$'s $100^3$, and open circles $128^3$. Error bar ranges for $U$ are about 1/30 the size of plotted points, between 1/4 and 1/10 for $\chi$, and 1/15 for $P$.}
          \label{fig4}
       \end{figure}
\begin{equation}
\chi = (<P^2>-<P>^2)N_p 
\end{equation}
and the corresponding Binder fourth-order cumulant 
\begin{equation}
U=(<P^4>-<P^2>^2)/(3<P^2>^2) .
\end{equation} 
Here $N_p$ is the number of plaquettes in the lattice. The correlation-length scaling hypothesis implies that these should collapse onto universal functions if scaled according to their respective exponents and plotted against the scaling variable 
\begin{equation}
x=(p-p_c)L^{1/\nu }
\end{equation}
where $L$ is the linear lattice size.  
Fig.~4abc shows the data for $U$, $\chi$ and $P$ as a function of the concentration of positive links $p$ for lattices of size $40^3$, $64^3$, $100^3$, and $128^3$.   All datapoints are from samples of 100,000 randomly generated lattices. One sees a crossing in $U$, similar to the case of a thermal phase transition. Here the crossing point marks the infinite-lattice percolation threshold.   Fig.~5ab shows the scaling collapse plots, where $\nu$, $\beta$, $\gamma$ and $p_c$ are adjusted to give the best collapse. Here the scaled $\chi$ is $\chi L^{-\gamma /\nu}$ and scaled $P$ is $PL^{\beta / \nu}$.   Although a good fit can be achieved using all four lattice sizes, a small systematic shift was seen in exponents toward typical percolation values when the $40^3$ data were omitted, suggesting a small correction-to-scaling effect of order the random error. For this fit there are 65 degrees of freedom overall and the fit to the three universal functions (in this case power laws) has $\chi ^2/$d.f$=0.77$.  The fit gives $\nu=0.872(4)$, $\beta /\nu =0.472(3)$, $\gamma / \nu  = 2.056(4)$ and $p_{c}=0.137317(5)$.  The latter agrees with that determined from finite-size shift above as well as with the threshold previously measured for NN+NNN site percolation, $p_{c}=0.1372(1)$\cite{nn+nnn}, which we believe to be equivalent to boundary percolation.  As far as we know exponents have not been previously measured for these cases.   The quantities $\gamma / \nu$ and $\beta /\nu$ should be related by the hyperscaling relation 
\begin{equation}
\gamma /\nu + 2\beta /\nu =d
\end{equation}
where $d$ is the spatial dimension.  Our values give, for the LHS, $3.001(7)$.
 For random site percolation a fairly recent high statistics study gives $\nu=0.8764(11)$ and $\beta /\nu =0.47705(15)$\cite{wang}. Comparing with our result leads to the conclusion that all of the exponents for random boundary percolation likely match those of ordinary site percolation.  
\begin{figure}[th!]\centering
                    \includegraphics[width=0.49\textwidth,  clip]{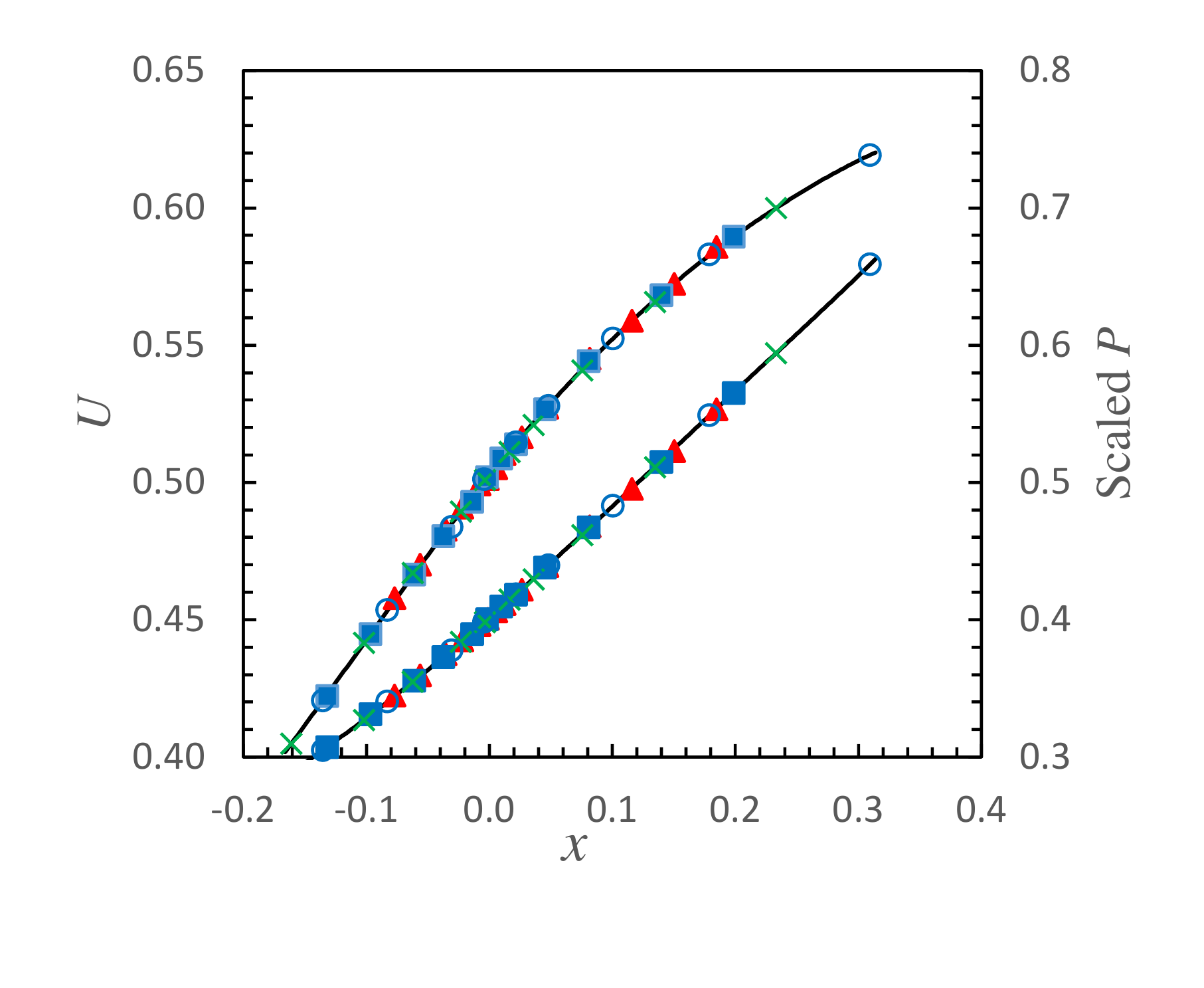}
                         \includegraphics[width=0.49\textwidth,  clip]{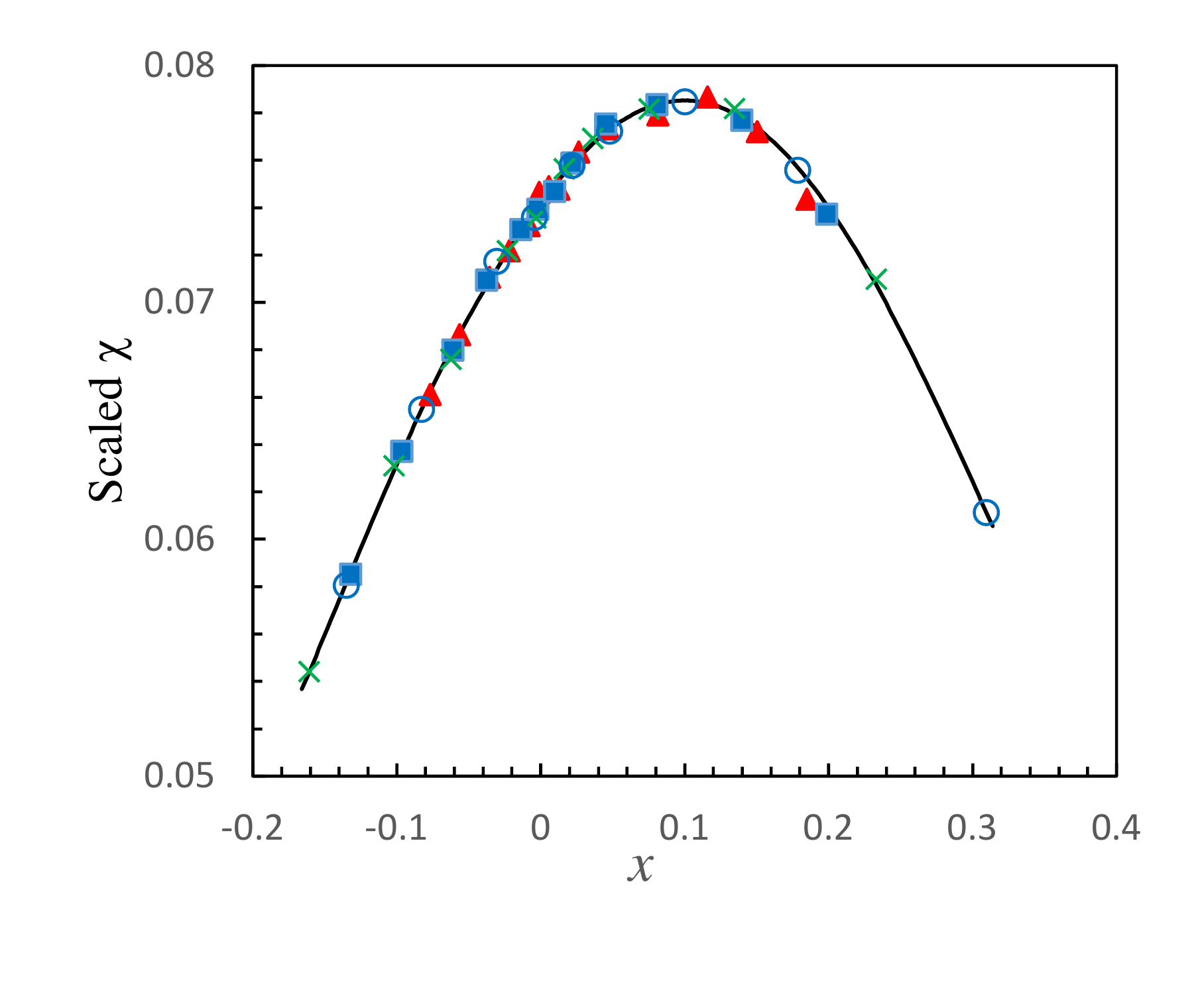}
                                  \caption{Scaling collapse plots for boundary percolation in the 3D random percolation model. Binder cumulant (left graph) and scaled largest-cluster fraction (a), and scaled susceptibility (b).}
          \label{fig5}
       \end{figure}

For random boundary percolation, finite-size shift and largest cluster scaling give consistent measurements of $\nu$. However that is not the case for boundary percolation in the 3D Ising model.  Figs.~6abc and 7ab analyze the largest cluster scaling for boundary percolation in the 3D Ising model in the same way as above. Of course now the abscissa is the Ising coupling strength $\kappa$.  The study was similar to the above but with 1,000,000 sweeps per point, sampled every 10, and 200,000 initial equilibration sweeps on $40^3$, $64^3$ and $100^3$ lattices.
Again we have a Binder cumulant crossing and excellent scaling collapse plots. These give $\kappa _c = 0.247925(6)$, $\nu = 0.867(5)$, $\beta /\nu = 0.465(5)$, and $\gamma /\nu = 2.068(20)$.  So there are no surprises here as these exponents are consistent with the random percolation values.  The fraction of minority sites at $\kappa _c$ is $0.1318(4)$,  about 4\% lower than for random percolation.  However, the finite-size-shift exponent, $\nu $, obtained above from the fit to Fig.~2b was $1.30(3)$ .  This is clearly incompatible with the percolation value just obtained from largest cluster scaling in the same system. This would seem to indicate that some other dynamics has taken over the scaling of the finite-size shift, driven by another correlation length which is becoming infinite at a different rate.  One possibility for how this could happen is if the percolation is linked to a thermal phase transition which has its own correlation length controlled by different dynamics.  This is a very curious behavior that invites further investigation because, as previously mentioned, it is quite unusual for a system to have two different correlation lengths.

\begin{figure}[thb!]\centering
                    \includegraphics[width=0.49\textwidth,  clip]{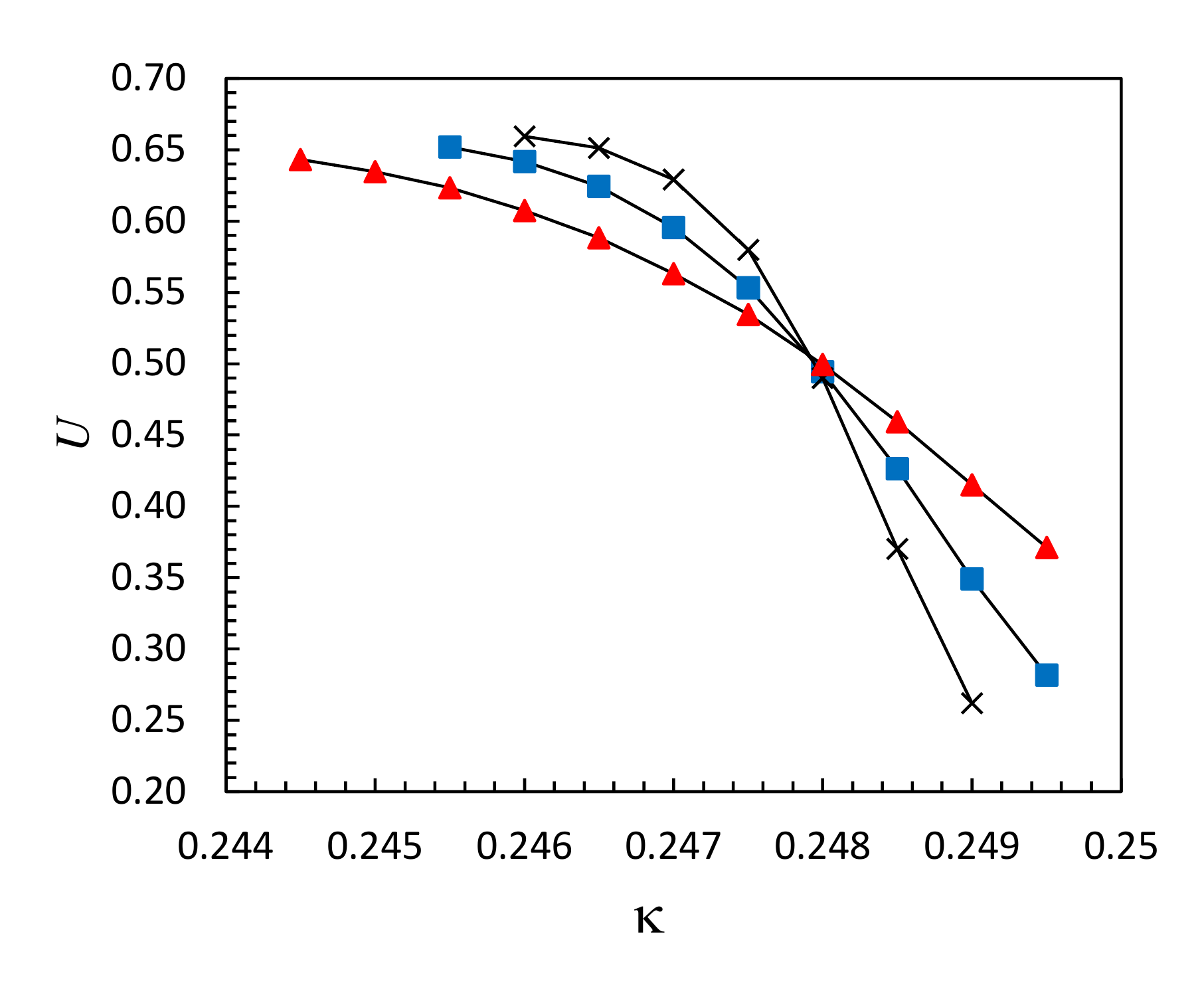}
                         \includegraphics[width=0.49\textwidth,  clip]{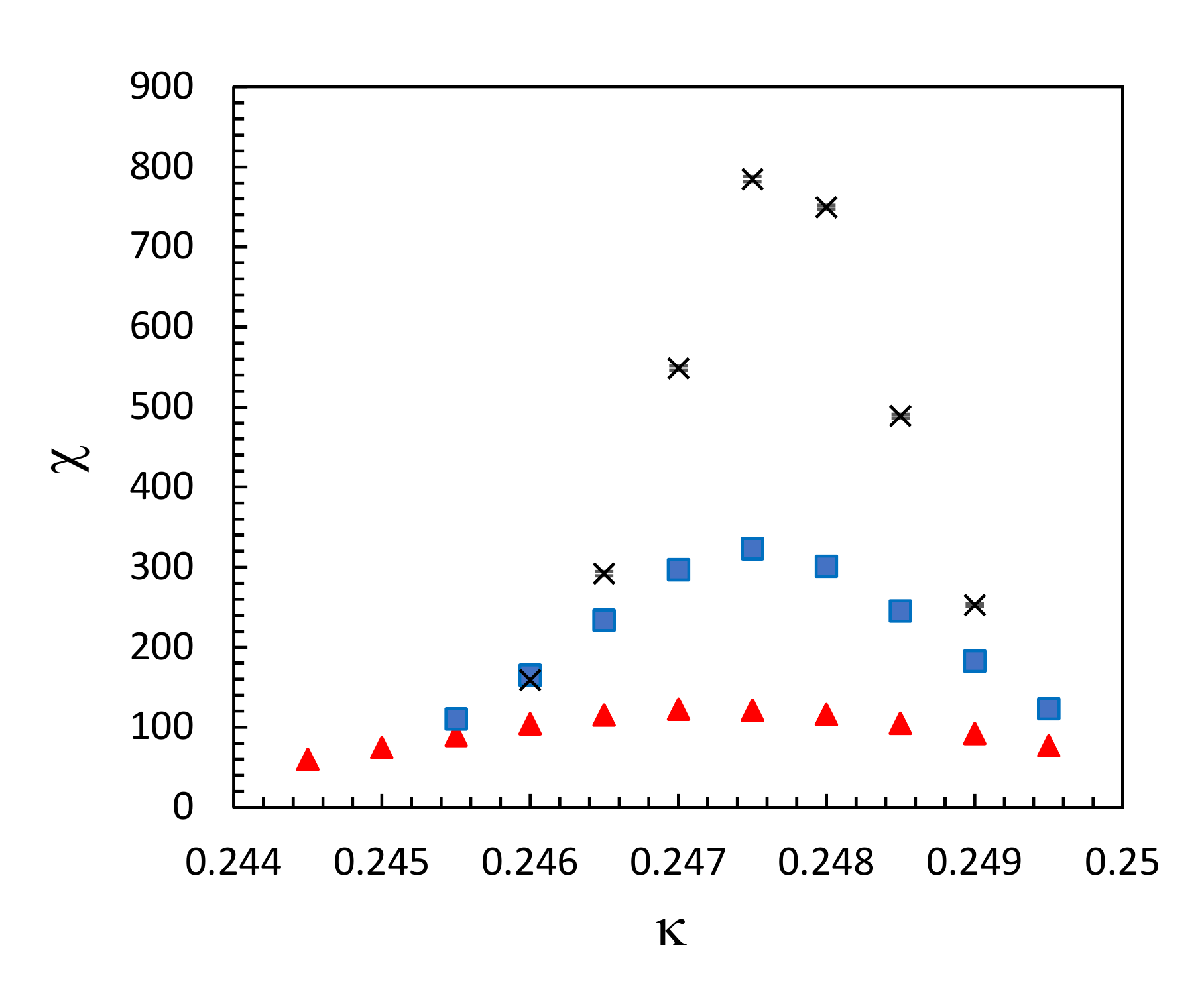}
                         \includegraphics[width=0.49\textwidth,  clip]{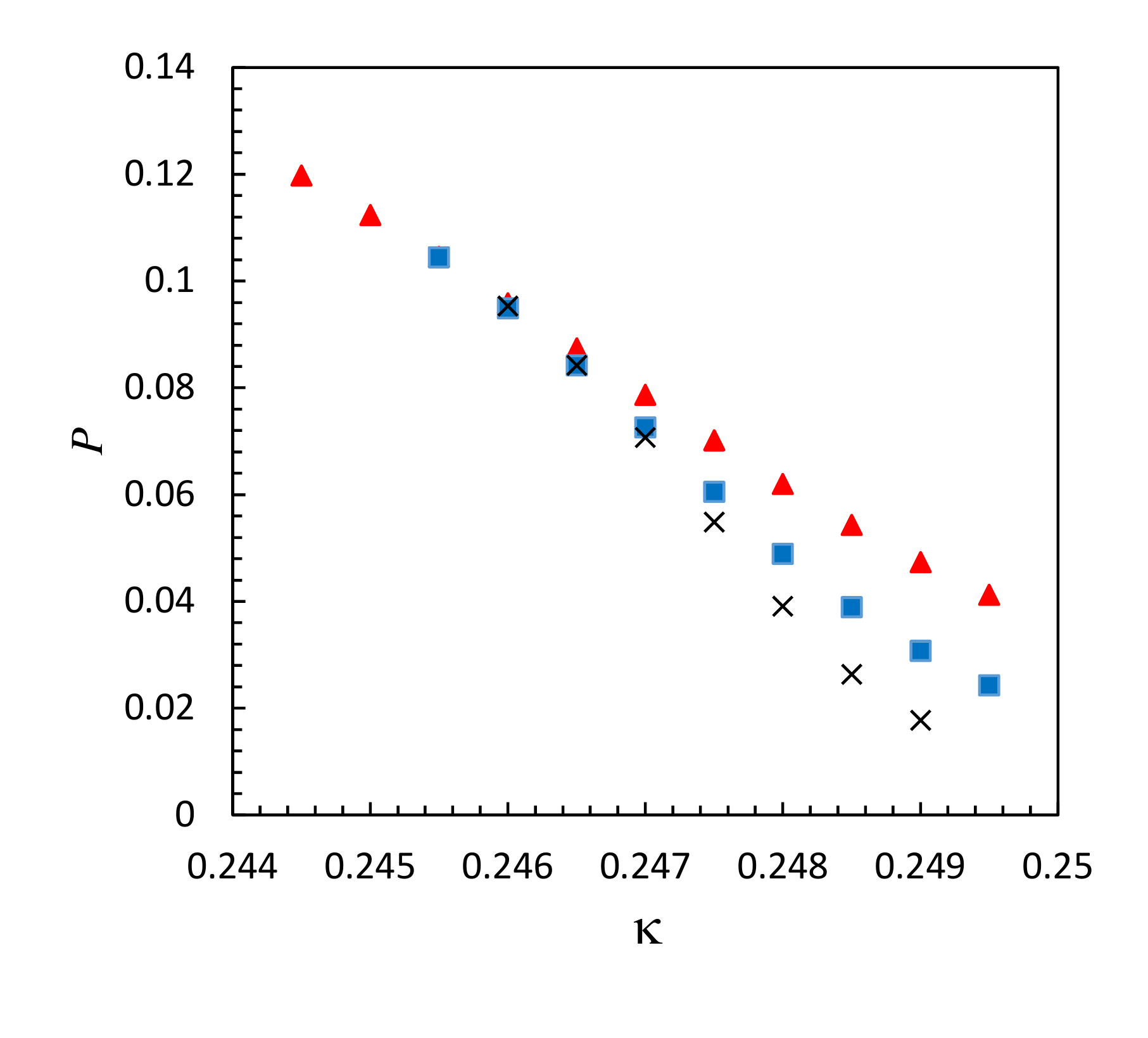}
                                  \caption{Boundary percolation study in the 3D Ising model.  Binder cumulant $U$ (a), susceptibility $\chi$ (b), and fraction of sites occupied by largest cluster $P$ (c) vs. coupling $\kappa$. Error bar ranges for $U$ are about 1/30 the size of plotted points, 1/20 for $P$, and between 1/5 and 1/20 for $\chi$.}
          \label{fig6}
       \end{figure}
\begin{figure}[thb!]\centering
                    \includegraphics[width=0.49\textwidth,  clip]{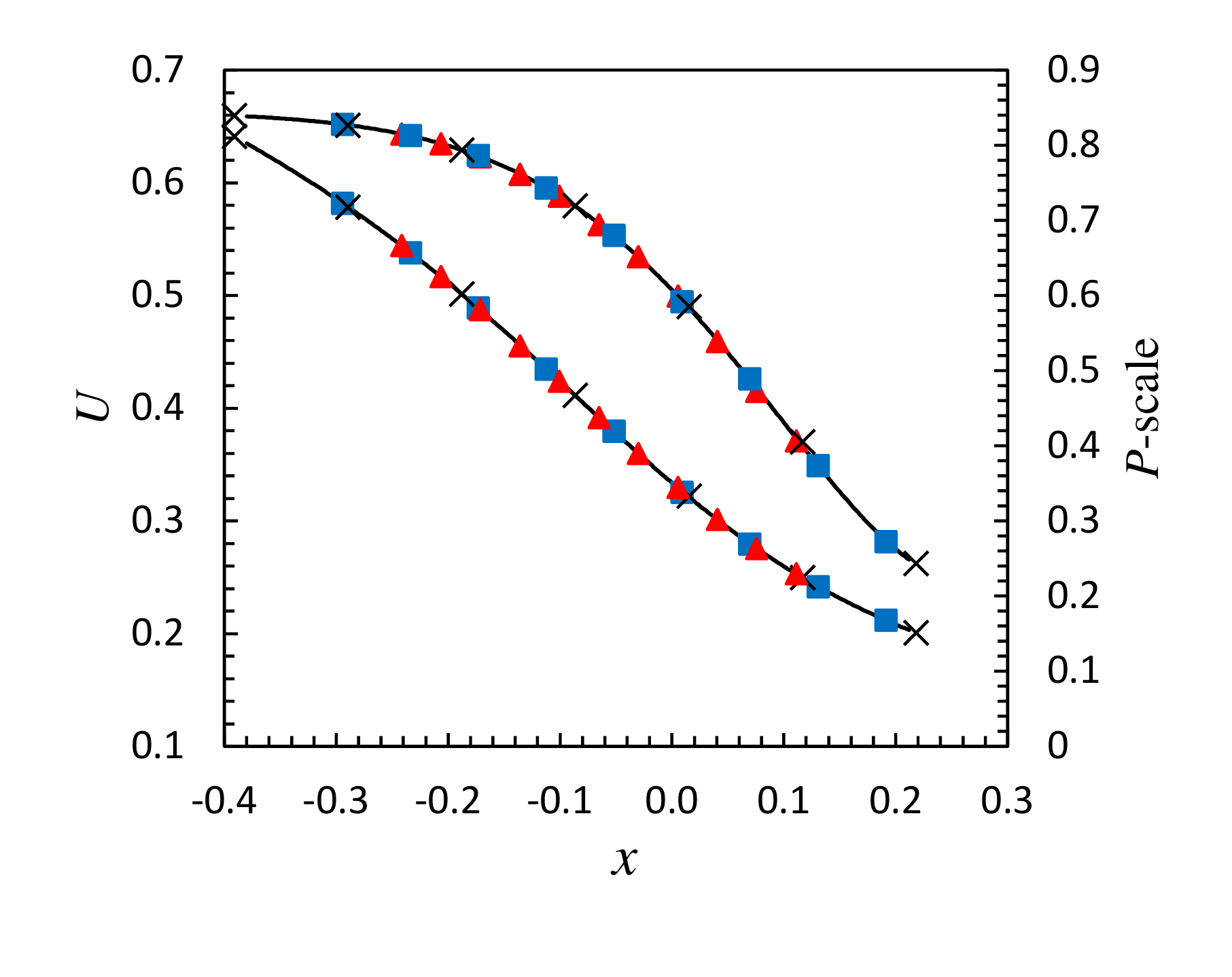}
                         \includegraphics[width=0.49\textwidth,  clip]{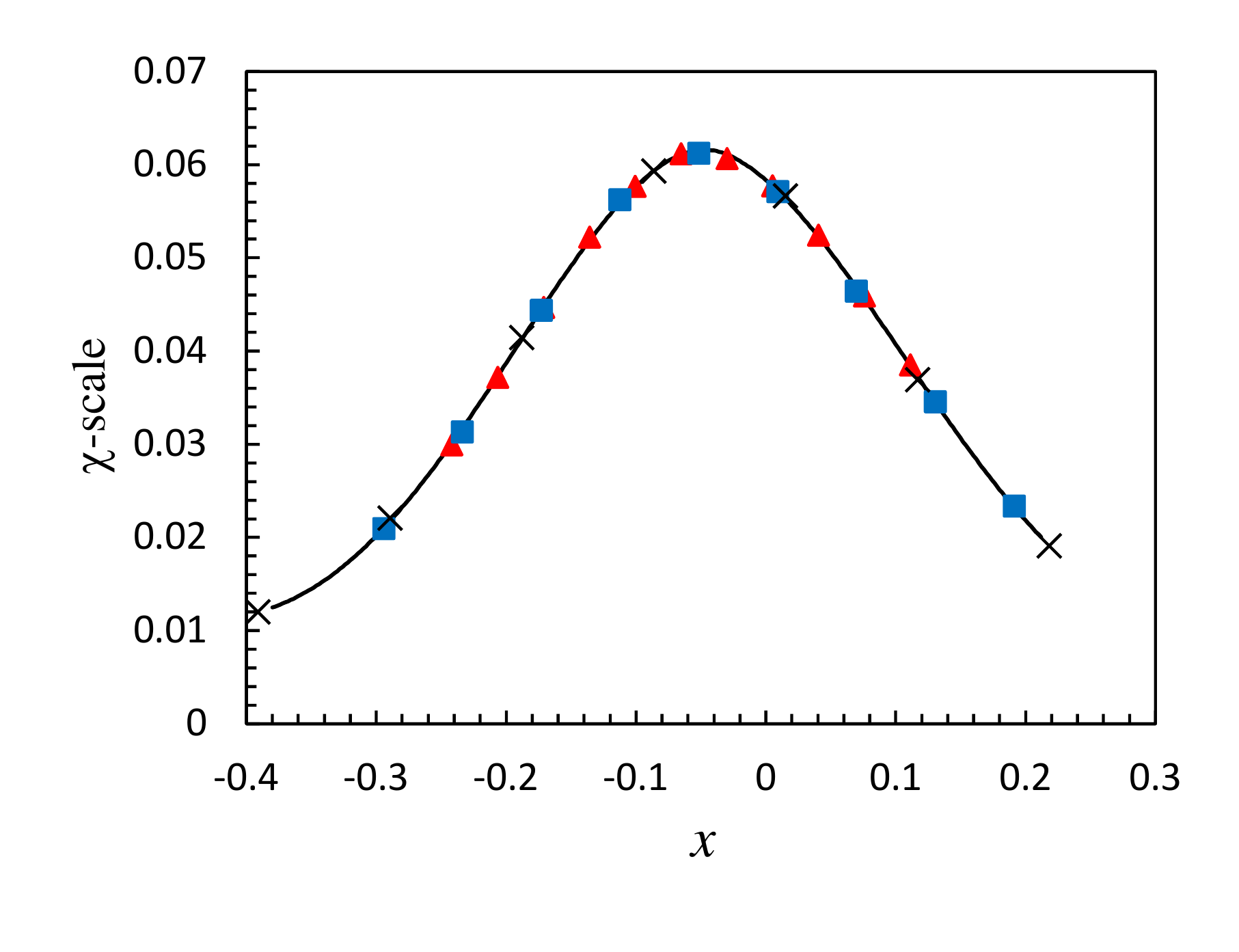}
                                  \caption{Scaling collapse graphs for percolation in the 3D Ising model.  Binder cumulant (left graph) and scaled largest-cluster fraction (a), and scaled susceptibility (b).}
          \label{fig7}
       \end{figure}

\section{Dual order parameter}
  As is well known, percolations are not necessarily coincident with phase transitions,
but sometimes are.  The situation is clearer if a symmetry-breaking order parameter
exists. In that case an energy singularity follows from the hyperscaling relation
$\alpha = 2-d\nu $, where $\alpha$ is the specific heat exponent, $d$ is the number
of dimensions and $\nu$ is the correlation length exponent associated with the 
order parameter near the symmetry-breaking phase transition. The spontaneous breaking of 
an exact symmetry is always associated with a mathematical singularity because the 
order parameter is exactly zero in the unbroken phase, and is non-zero in the
broken phase\cite{landau}. A function which is zero over a range of values can only become
non-zero at a point of non-analyticity. 

In order to build further evidence of a phase transition at the point of boundary
percolation, one can examine the dual theory, the three-dimensional gauge Ising model.
This has action 
\begin{equation}
S=-\beta\sum _p U_p
\end{equation}
where $U_p$ is the product of four gauge fields $U_{\mu ijk}$ around an elementary plaquette. The $U_{\mu ijk}$ exist on links with $\mu$ a direction index and $ijk$ the site address.
The duality relation maps the coupling of the Ising model $\kappa$
to $\beta$ of the dual gauge theory, $\beta = -0.5\ln(\tanh(\kappa))$\cite{duality}. 
The ordered phase of the spin theory maps to the disordered (confining) phase
of the gauge theory. Generally it is not considered that there is a local
symmetry-breaking order-parameter in gauge theories, because Elitzur's 
theorem\cite{elitzur} does not allow a local symmetry to break spontaneously. 
However, if one transforms configurations to Coulomb gauge then a symmetry-breaking 
order parameter may be defined, for which the remnant symmetry breaks in the deconfined
phase\cite{coulomb-confinement}. The Coulomb gauge transformation seeks to maximize the 
number of positive links in the one and two directions, ignoring the third direction
links. This leaves a remnant layered Z2 symmetry. Two-dimensional global symmetry operations applied to 
single 1-2 layers do not alter the one and two direction links on which Coulomb gauge is
defined, but flip all third direction links attached to the layer.  For fixed one
and two direction links the third direction links have mostly ferromagnetic interactions
from plaquettes with two positive one or two direction links, especially at high $\beta$.
If one takes the third direction links in each separate layer as order parameters,
it is found that these magnetize exactly at the dual-reflection of the 3-d Ising
critical point\cite{previous}. The deconfined phase is magnetized and the confined
phase is not.  The dual reflection of the boundary-percolation point lies in the 
confined phase, ie. the non-magnetized phase of the gauge theory. If there is a
symmetry-breaking phase transition here it must be a spin-glass transition, which is
a symmetry-breaking transition within the unmagnetized phase. A spin glass has a hidden 
pattern of order which does not result in an overall magnetization. To search
for such a transition we used a two-real-replica approach\cite{2rr}. A second set of third-
direction pointing links is equilibrated to a fixed pattern of one and two direction
links from the main simulation. This is similar to the initial equilibration
for any Monte-Carlo simulation. Then the order parameter is defined as
\begin{equation}
q_k =\sum _{i,j}R_{3ijk}U_{3ijk}  .
\end{equation}
Here $R_{3ijk}$ is the replica third-direction link at site $ijk$ and $U_{3ijk}$ is the 
original one. Note there is a separate $q_k$ for each 2D layer, because the symmetry being broken is only global in two directions but still local in the third direction.  As is usual one needs to take the absolute value of the order parameter due to tunneling on the finite lattices.  We also choose to take the square root of the order parameter since it is the product
of two spins, but this is not absolutely necessary.  Thus the average spin-glass magnetization to be analyzed is 
\begin{equation}
M\equiv <\sqrt{|q_k|}>
\end{equation}
where the average is both over gauge configurations as well as third direction fixed 2D layers in each gauge configuration. The order parameter $M$ will become non-zero in a phase with either spin-glass order or
ferromagnetic order. Spin-glass order is symmetry breaking because the symmetry operation
applied only to the original U's but not the replicas will invert the order parameter.
Another way to say this is that tunneling configurations within the replica or original,
where half of the lattice is flipped, do not exist in the spin-glass phase in the 
thermodynamic limit.  For systems without a spin-glass phase this order parameter
will simply turn on at the normal ferromagnetic transition (for instance, this is the case 
for Landau-gauge Higgs phase transitions in the combined Ising gauge-Higgs theory\cite{previous}).
Note also that although its original motivation was from Coulomb gauge, $q_k$ itself is
gauge invariant, so it is no longer necessary to fix the gauge.  As detailed below, we indeed find a
phase transition in $M$ away from the ferromagnetic transition indicating the presence
of a spin-glass phase.  Here we can use all of the finite-size scaling 
techniques which have been developed
for studying symmetry-breaking phase transitions with a local order parameter.
\begin{figure}[th!]\centering
                    \includegraphics[width=0.49\textwidth,  clip]{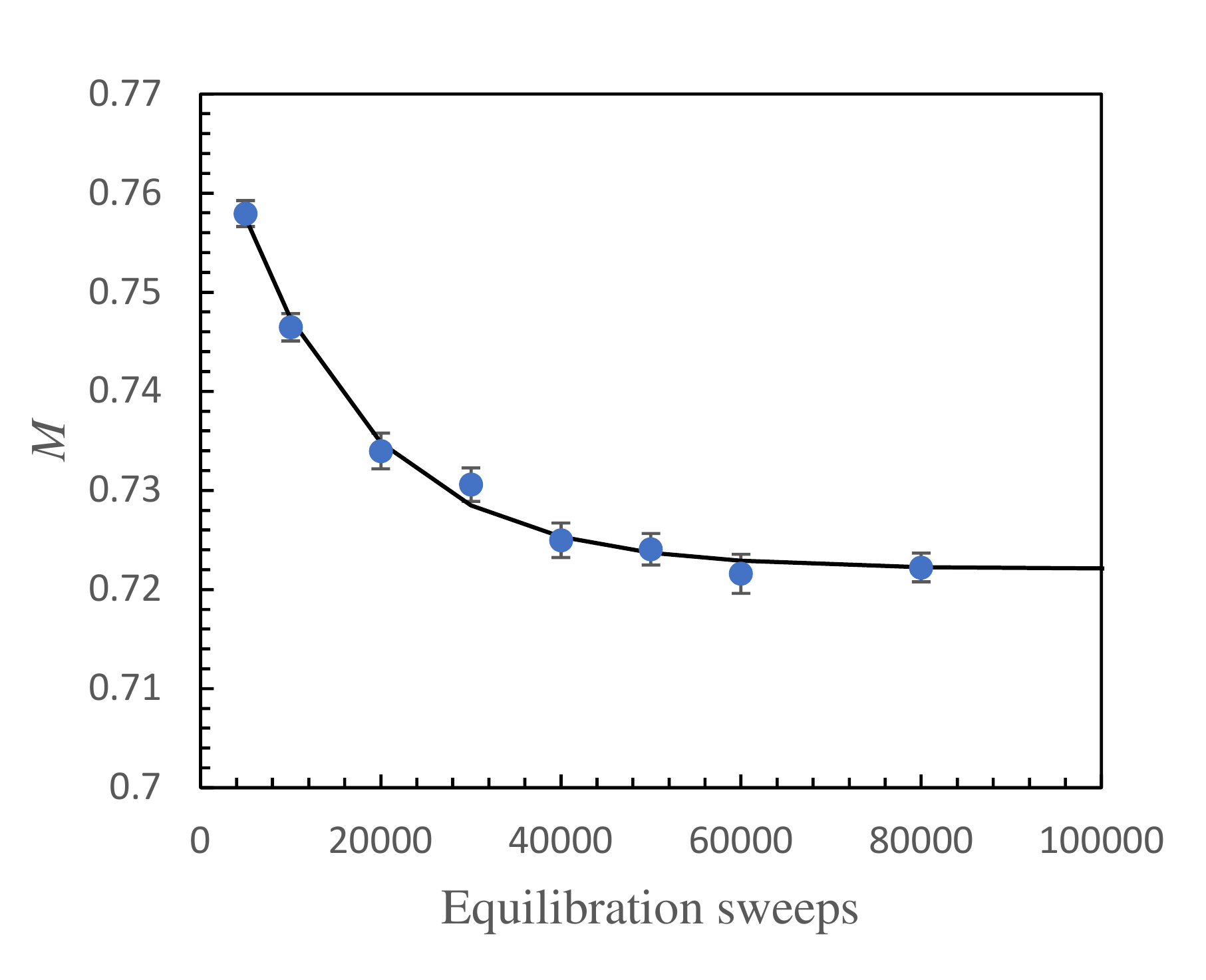}
                   \includegraphics[width=0.49\textwidth,  clip]{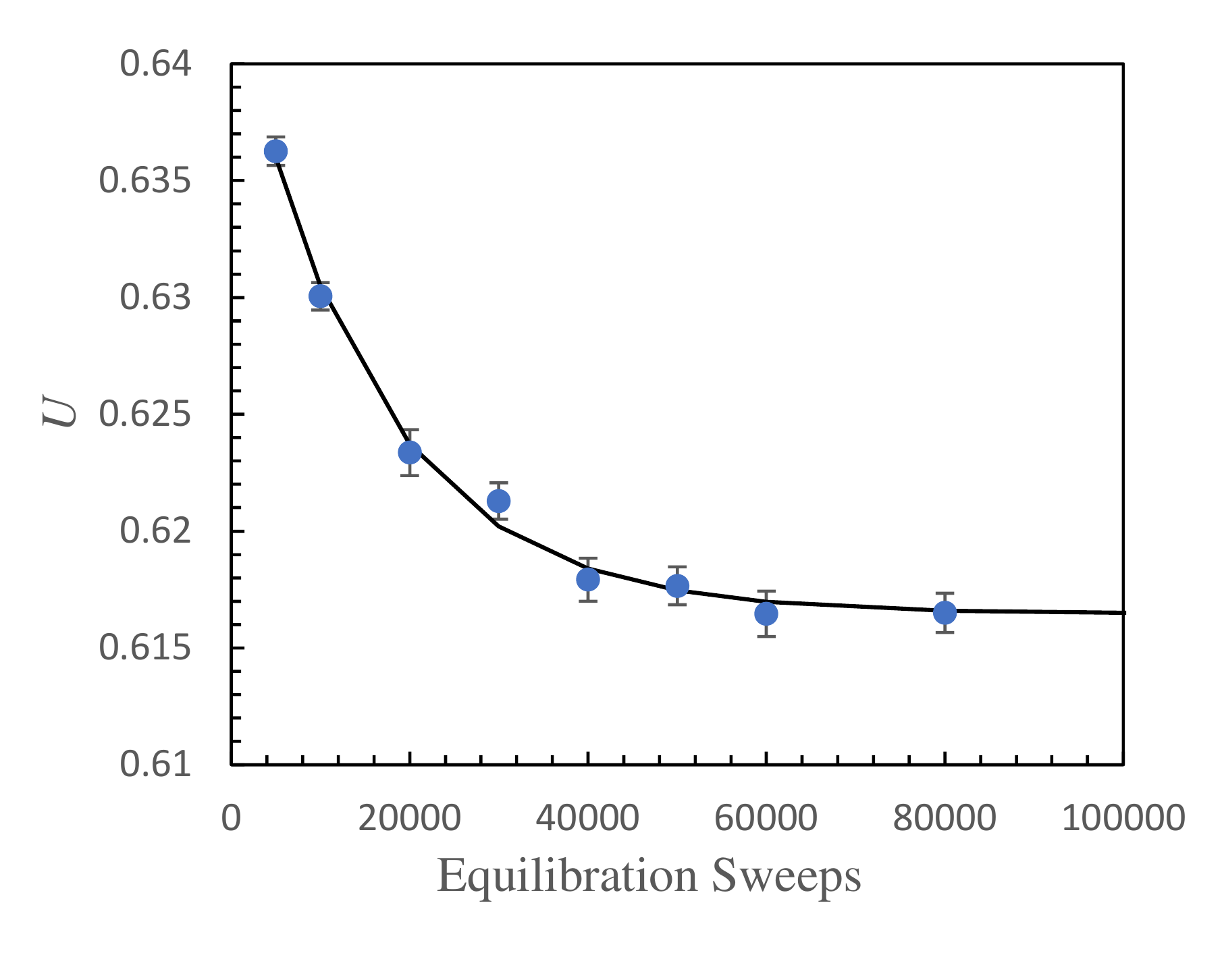}
                                                  \caption{Equilibration of spin-glass order parameter and Binder cumulant on a $30^3$ lattice.}
          \label{fig8}
       \end{figure}

Before studying the spin-glass order parameter $M$ in Monte Carlo simulations, one must first perform an equilibration study to determine how long the replica must be equilibrated to obtain a truly independent configuration.  One simply simulates at many different equilibration sweep values and watches the measured quantities approach constant values exponentially.  We then picked equilibration amounts that insure systematic errors are less than 25\% of random errors in the quantities measured.  Detailed studies were made at gauge coupling $\beta =0.705$, near the suspected critical point, for both the $30^3$ and $50^3$ lattices.  The equilibration value for the $40^3$ lattice was determined from these and the volume scaling suggested by them.  Fig.~8ab shows the equilibration of magnetization (order parameter) and its Binder cumulant for the $30^3$ lattice.  Other quantities were similar.  The exponential fits give an equilibration time constant of 14,000 sweeps.  By equilibrating with 105,000 sweeps systematic errors are brought to less than 25\% of random in the planned simulations.  For $50^3$ this value was a bit surprisingly high at 700,000 sweeps.  We used 190,000 sweeps for the intermediate $40^3$ case.  These high equilibration values indicate the standard heat bath Monte Carlo algorithm is not working particularly well here, but it still gives good results if one is patient.  The high number of sweeps to equilibrate are due to the fact that 2/3 of the links, those lying in the 1 and 2 directions are being held fixed, which erects more barriers that a simulation where all links participate.
\begin{figure}[thb!]\centering
                    \includegraphics[width=0.49\textwidth,  clip]{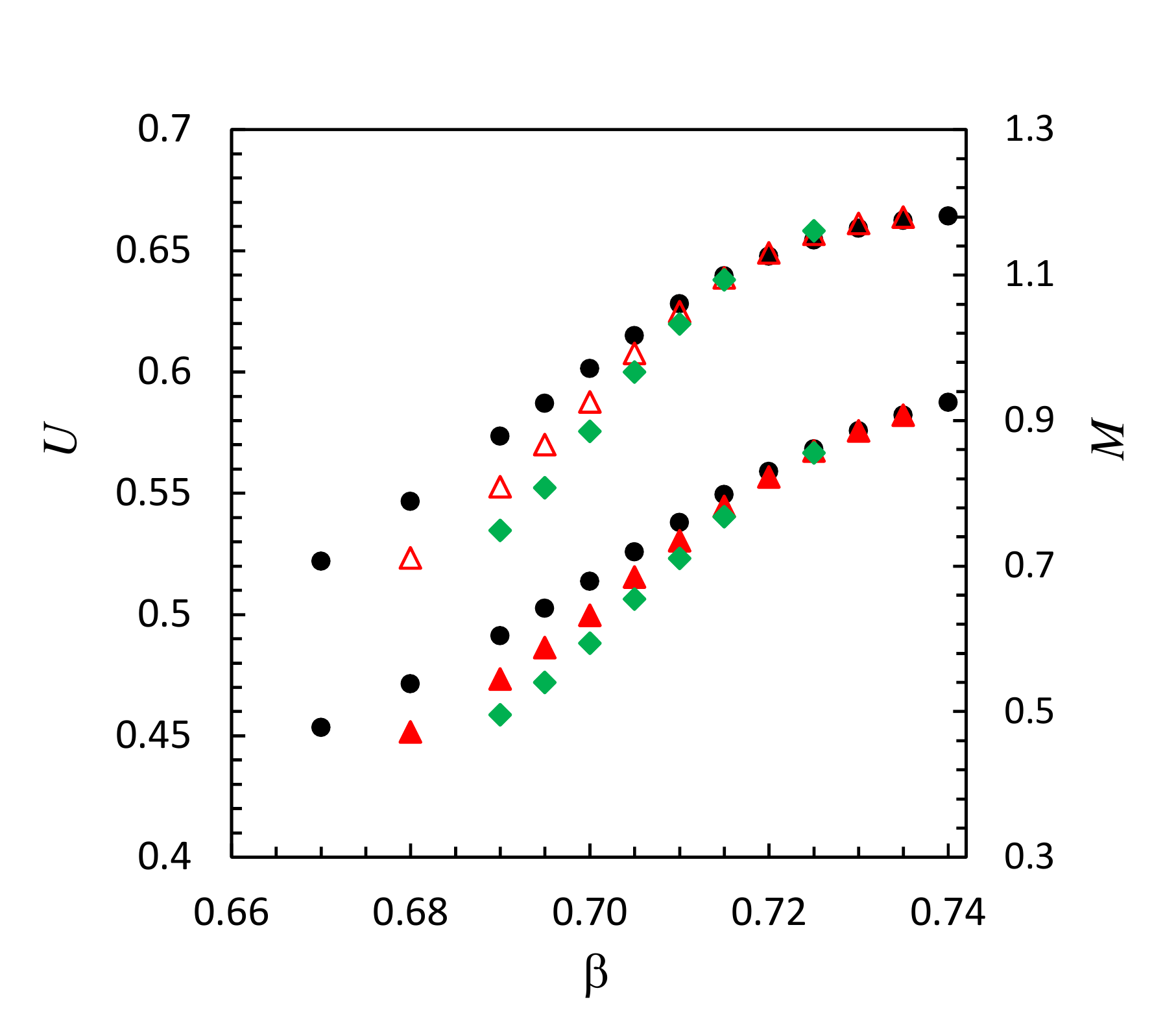}
                         \includegraphics[width=0.50\textwidth,  clip]{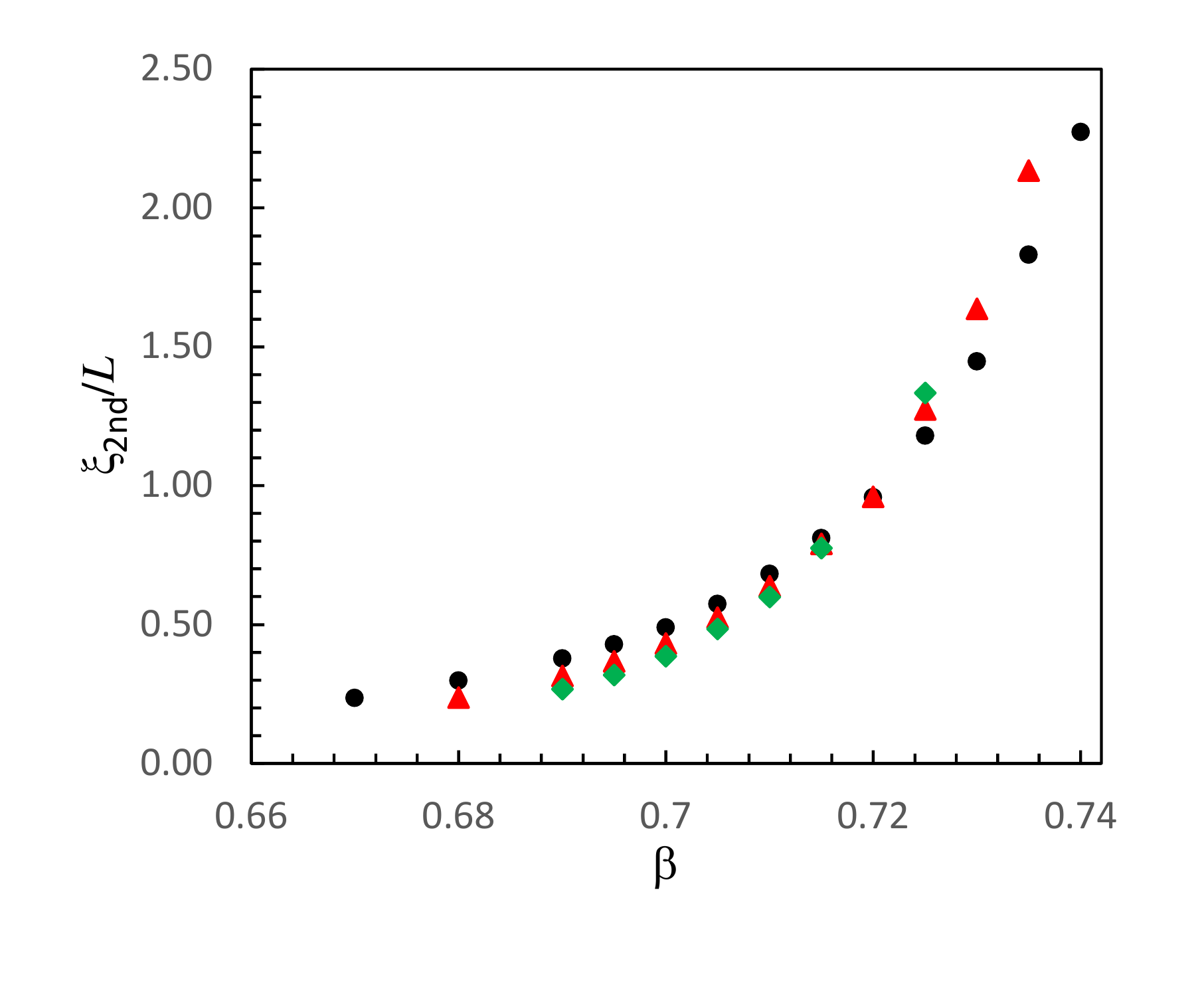}
                         \includegraphics[width=0.49\textwidth,  clip]{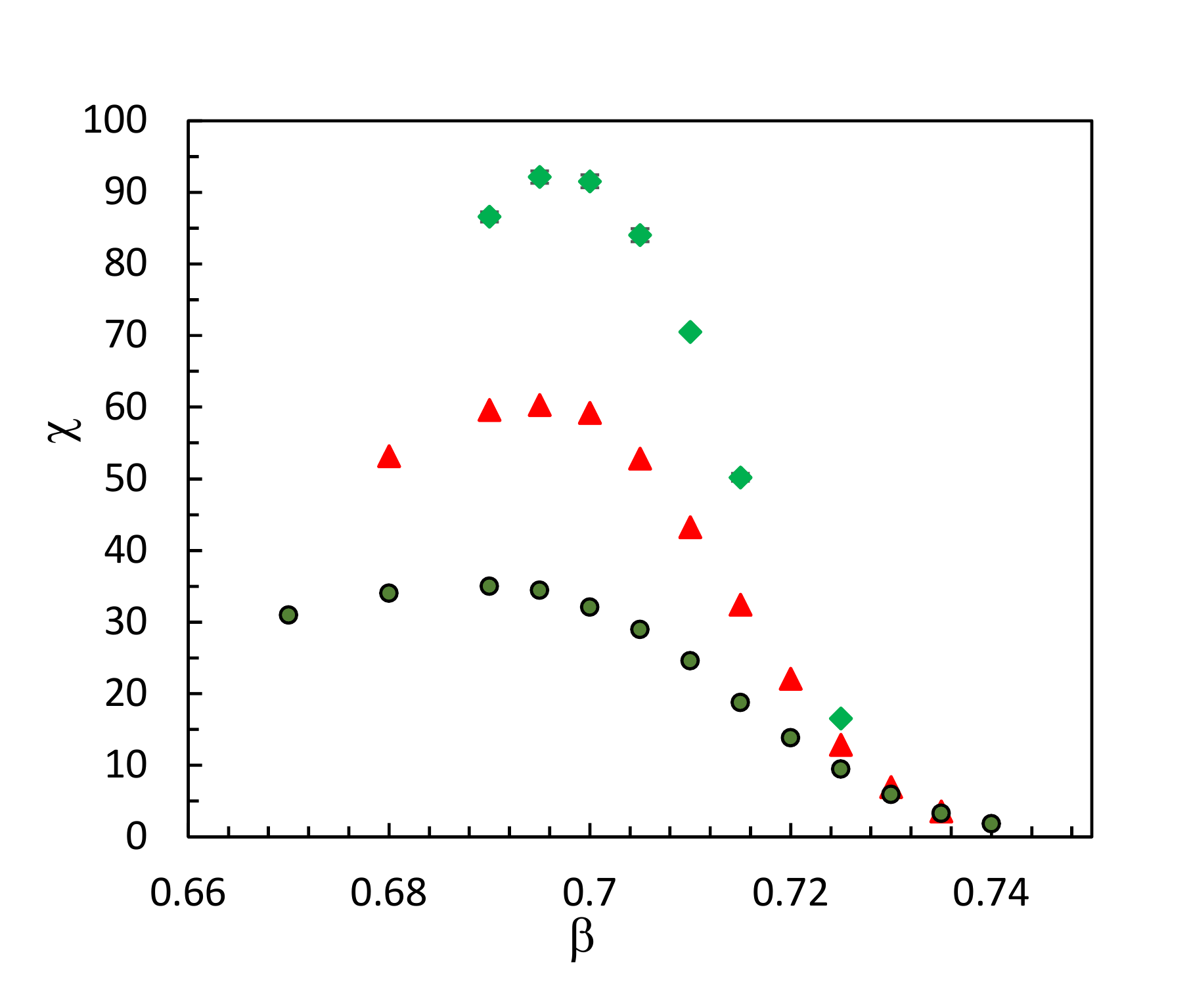}
                                  \caption{Binder cumulant (left graph) and magnetization (a), $\xi _{\rm 2nd}/L$ (b), and susceptibility (c) for the spin-glass order parameter. Error bar spreads for $U$ and $M$ are about 1/15 the size of plotted points, and 1/2 to 1/5 for $\xi _{\rm 2nd}/L$ and $\chi$.}
          \label{fig9}
       \end{figure}

All simulations had 100 ordinary Monte Carlo sweeps between each measurement of the order parameter to reduce correlations.  There were 1000 measurements for each $30^3$ and $40^3$ lattices and 500 for $50^3$. Initial equilibration was 200,000 sweeps.  Error bars were determined from binned fluctuations.
Fig.~9abc shows the Binder cumulant $U$, order parameter $M$, susceptibility $\chi$, and second moment correlation length\cite{smcl} divided by lattice size, $\xi _{\rm 2nd}/L$, for the three lattices.  The latter, as well as the Binder cumulant, should cross near the infinite lattice transition point (to determine this precisely one must consider corrections to scaling which we do not do here).  One can see a well-defined crossing in both near $\beta _c = 0.715$.  The crossings are well established. For instance
the $50^3$ value exceeds the $30^3$ value at $\beta=0.725$ by $15\sigma$ for $U$ and $10\sigma$ for $\xi _{\rm 2nd}/L$, and points above this have similar significances. The opposite order in the low $\beta$ region is never in doubt.
Indeed, here points here are separated by even larger amounts, exceeding $30\sigma$. Scaling collapse plots are shown in Fig.~10abc. The overall fit has $75$ degrees of freedom and has a $\chi ^2 /$d.f.$= 1.48$ .  This fit gives $\beta _c=0.7174(3)$, $\nu = 1.27(3)$, $\beta /\nu = 0.058(2)$, and $\gamma /\nu = 1.86(2)$. Checking hyperscaling on the latter give $d_{\rm eff}=\gamma /\nu + 2\beta /\nu =1.97(2)$. Because the order parameter is defined on 2-d layers, the expected value is 2.  The dual reflection of the boundary percolation point of the 3D Ising model itself is $-0.5\ln \tanh (0.247925)=0.70741(1)$.  This is close to the $\beta _c$ here, but certainly not an exact match, and not within statistical errors.  However there could be a systematic error present from corrections to scaling. Looking at the $U$ crossing (Fig.~9a), it is plausible that the crossing on larger lattices could shift to this point.  Corrections to scaling can give a slightly shifting crossing with increasing lattice size.  There is also the possibility of a residual systematic error from insufficient equilibration. Although we have tried to limit this to 25\% of the random error it could still have an effect.  The fact that the $\nu$ seen here and the $\nu$ from the coupling-shift of the percolation transition, $1.30(3)$ agree within $1\sigma$ strongly supports these being dual-manifestations of the same transition.
\begin{figure}[thb!]\centering
                    \includegraphics[width=0.52\textwidth,  clip]{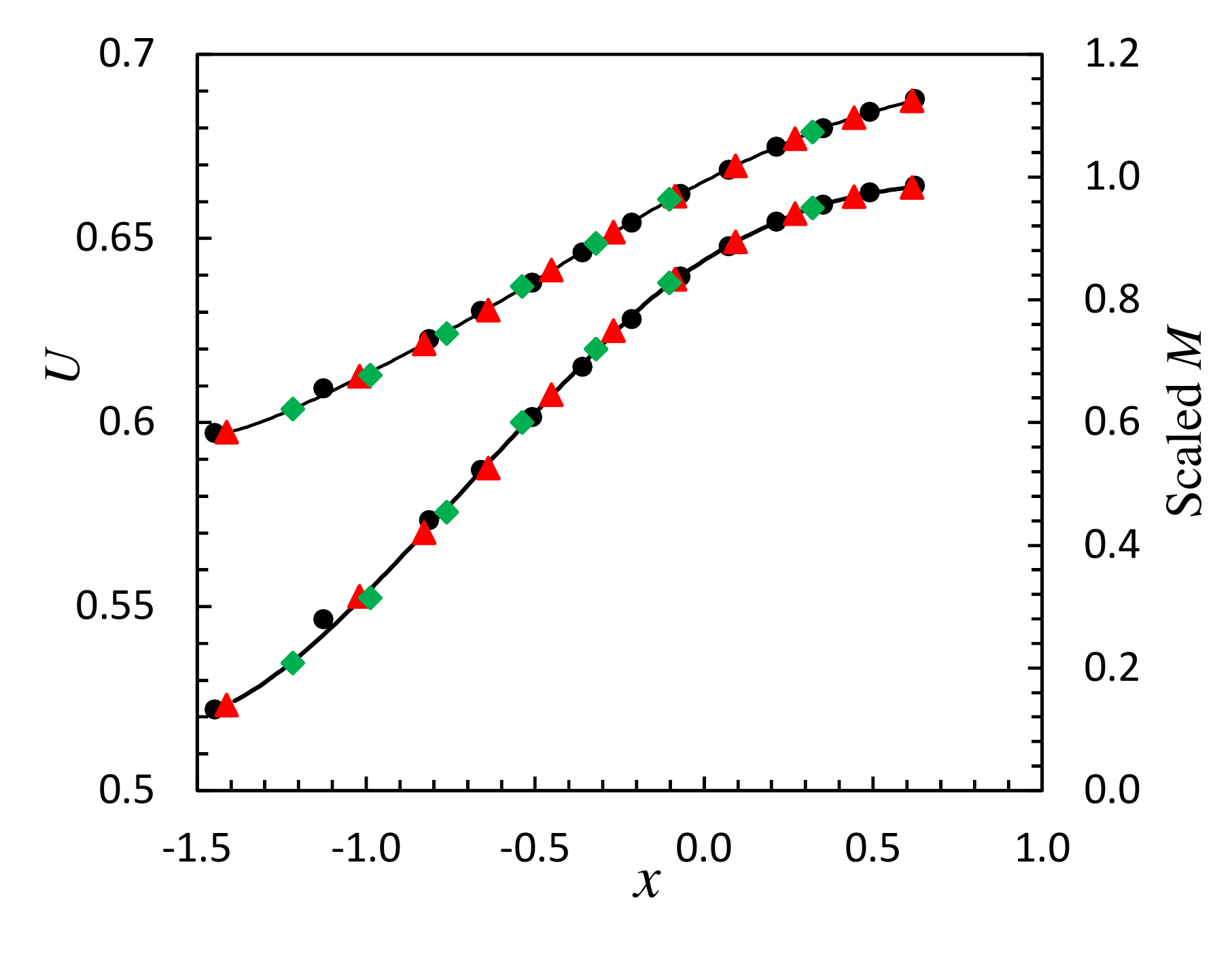}
                         \includegraphics[width=0.47\textwidth,  clip]{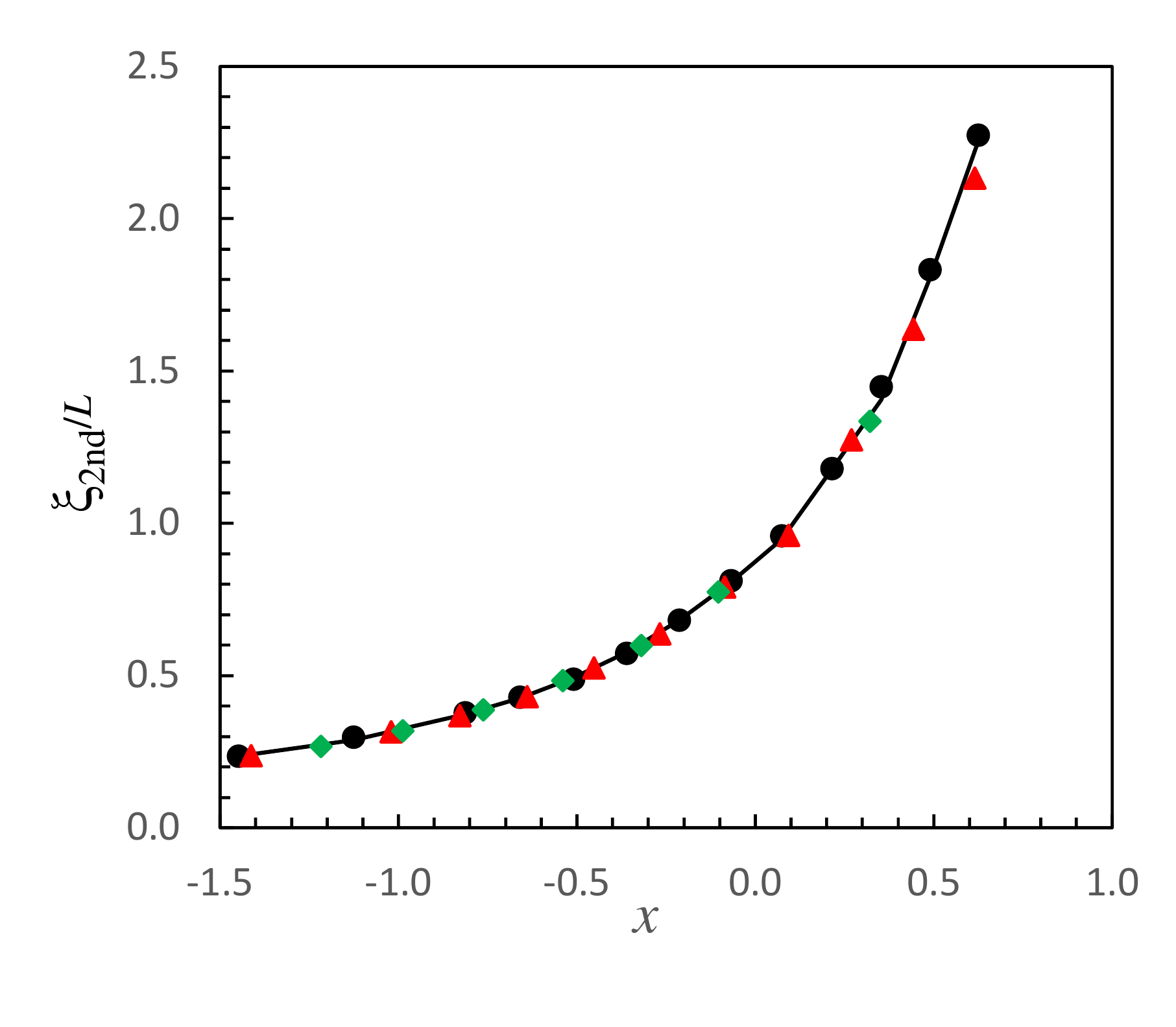}
                         \includegraphics[width=0.49\textwidth,  clip]{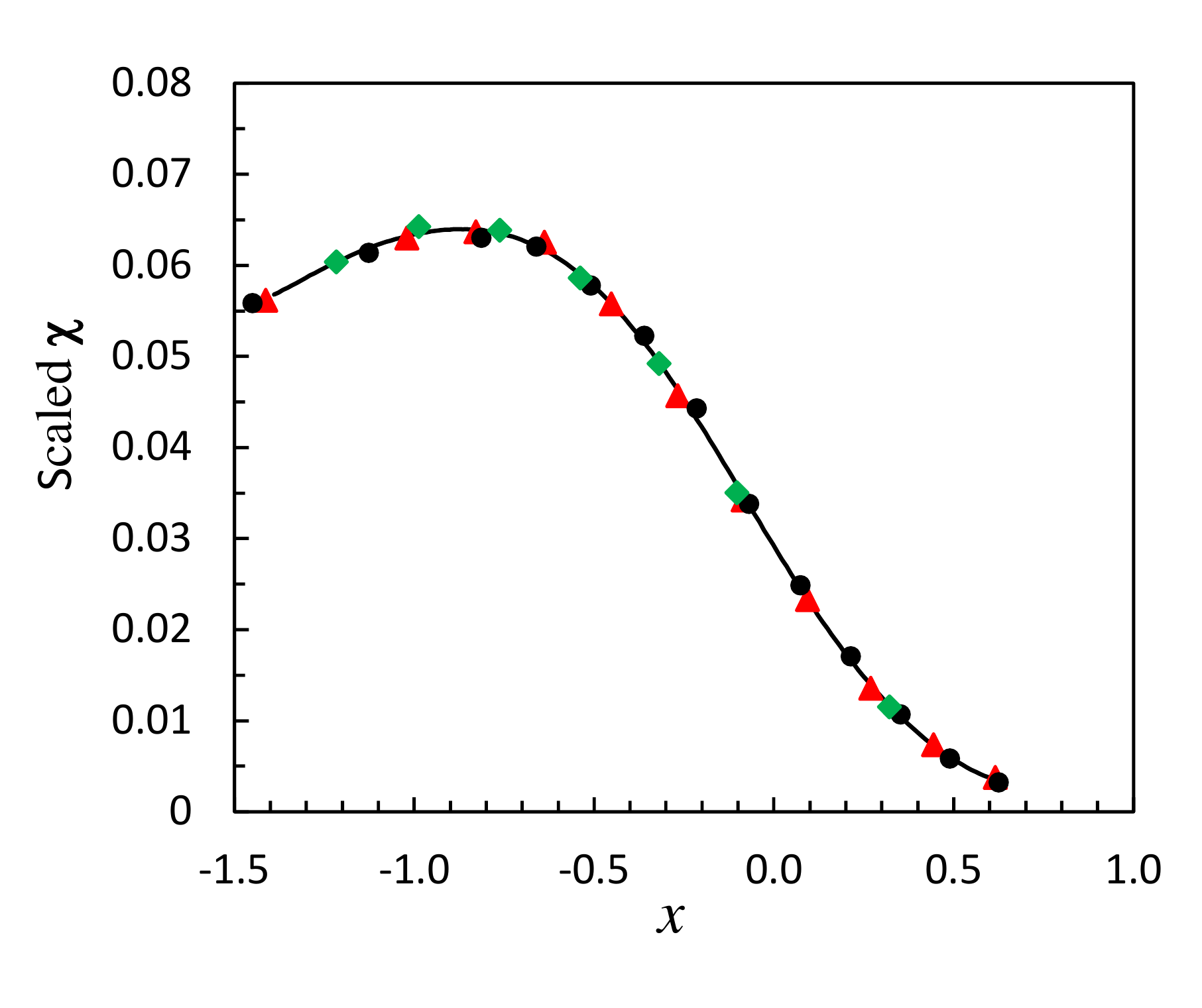}
                                  \caption{Scaling collapse plots for Binder cumulant (left graph) and scaled magnetization (a), $\xi _{\rm 2nd}/L$ (b), and scaled susceptibility (c), for the spin-glass order parameter.}
          \label{fig10}
       \end{figure}

\section{Energy moments}
  Since the spin-glass transition in the dual theory is symmetry breaking, Landau theory connects this to a thermal phase transition through the hyperscaling relation 
\begin{equation}
\alpha = 2-d\nu .
\end{equation}
Here $\alpha$ is the specific heat exponent. At the critical point the expected behavior of the specific heat is $|T-T_c|^{-\alpha }$. For $\nu= 1.3$, $\alpha = -1.9$.  This means that the specific heat does not have an infinite singularity, however it does have a finite singularity.  Unfortunately, when rounded by a finite lattice size, these are difficult to spot using
finite-size scaling. Nevertheless one can still try to fit to a fractional power, and in some cases more importantly, a different coefficient on each side of the transition.  However, the energy moments also have non-singular terms. This makes fitting them more difficult than quantities based on the order parameter which are purely singular. The non-singular part is expected to vary slowly through the critical region. For this reason it affects higher moments less, and there is a good chance these can be fit without a non-singular part other than perhaps a constant.  This simplifies fitting to the expected critical behavior.  A study of energy moments of the 3D Ising model itself was performed.  This study had approximately $7\times 10^9$ sweeps at each coupling for the $30^3$ lattice and $2\times 10^9$ for the $40^3$ and $50^3$, with measurements performed every other sweep.  With these statistics, rather precise data can be obtained on the third cumulant (third central moment), defined as $<(E-\bar{E})^3>(3L^3 )^2$. It is this combination that corresponds to the derivative of the specific heat.  The third cumulant is expected to scale as $|\kappa-\kappa_c|^{-\alpha -1}$, which is still a non-infinite singularity.  This quantity is shown in Fig.~11, along with a fit to the expected critical behavior, but leaving $\alpha$ and $\kappa _c$ as free parameters. The fit also allows for a different coefficient on the two sides of the transition. The result is 
$\kappa _c=0.2477(2)$, and $-\alpha -1=0.967(12)$.  The coefficient ratio below and above the critical point is $1.287(25)$.  The predicted $\nu$ from this $\alpha $ is $\nu=(-\alpha +2)/3=1.322(4)$.  The critical point agrees well with that extracted from percolation($0.24781(4)$, and the exponent $\nu$ also agrees with those from both percolation finite-size shift ($1.30(3)$) and the spin glass transition in the dual gauge theory($1.27(3)$).  Even though the singularity is non-infinite, it can still be seen from this fit.  It is important to remember that these functions are singular in two ways - the fractional power and the jump in coefficient.  So even if the power were to end up being exactly unity, that would not erase the singularity due to the fairly large coefficient jump, verified to $11.5 \sigma$, which can be seen in the change of slope.

\begin{figure}[th!]\centering
                    \includegraphics[width=0.60\textwidth,  clip]{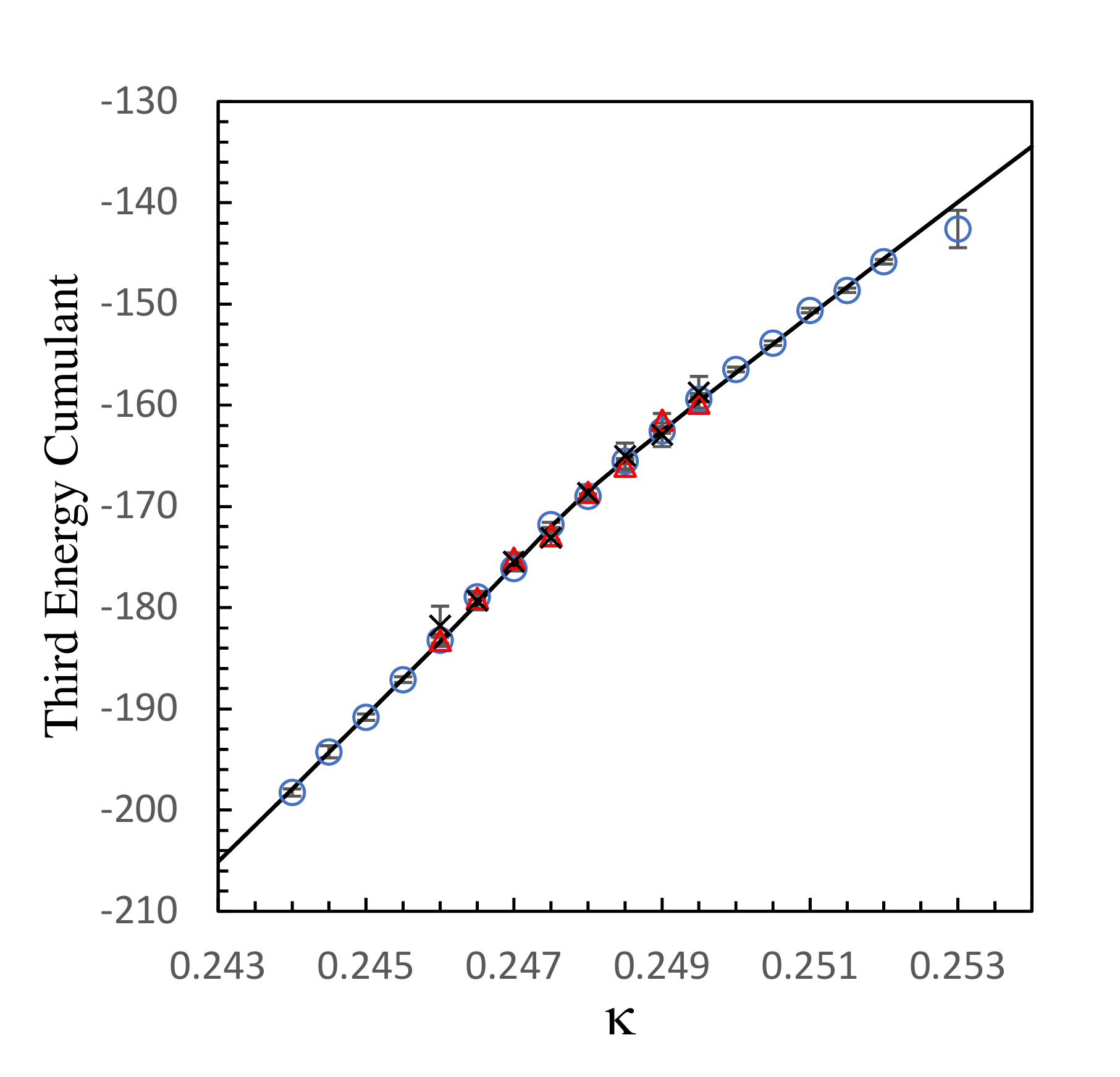}
                                                  \caption{Third order energy cumulant, with fit to critical behavior.  Here open circles are $30^3$, open triangles $40^3$, and $\times$ $50^3$.}
          \label{fig11}
\end{figure}
Higher moments were also measured, but even with these high statistics were somewhat of a disappointment due to fairly large statistical errors. Fig.~12 shows the fourth cumulant,
\begin{equation}
(<(E-\bar{E})^4>-3<(E-\bar{E})^2>^2) (3L^3 )^3 .
\end{equation}
This combination of moments tracks the third derivative of the internal energy with respect to $\kappa$.  Also shown is a numerical derivative of the fit function to the third cumulant on a parameter spacing $1/4$ of that used for the simulations.  This was done instead of an exact derivative to simulate finite-lattice rounding, so not an exact prediction of the expected behavior, but one which should be good away from the critical point.  The main effect of the shift in slope in the third cumulant which translates to a shift in level here can be seen.  In principle some finite-size effect could be seen in this quantity since it diverges with a very small exponent, but the expected ratio in peak heights between $30^3$ and $40^3$ is only $(4/3)^{((\alpha +2)/\nu)}=1.02$, much smaller than our statistical errors.  A larger effect is predicted for the fifth cumulant ($1.28$), but the errors there are magnified even more.  This figure is shown primarily to illustrate how much further one would have to go in statistics to see an infinite singularity in a high moment.
Our program, which was run for about 24 processor-years on PC's, does not employ multi-spin coding.  Perhaps a study that did or used specialized hardware could see these effects.
\begin{figure}[th!]\centering
                    \includegraphics[width=0.60\textwidth,  clip]{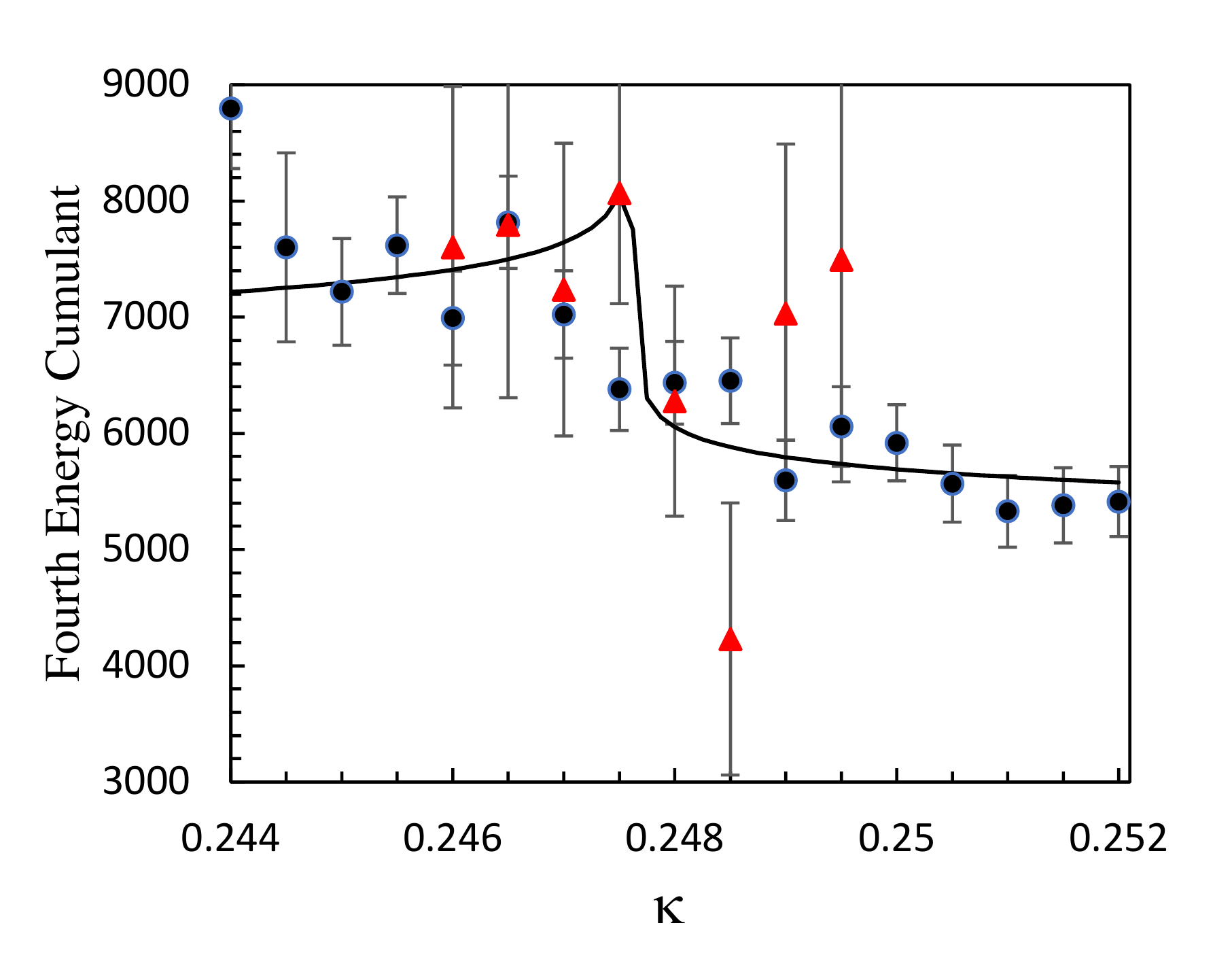}
                                                  \caption{Fourth order energy cumulant for $30^3$ and $40^3$ lattices.  Line is a plausible rounded critical behavior based on third cumulant fit (see text).}
          \label{fig12}
\end{figure}

\section{Conclusion}
In this paper evidence has been given for a new high-order phase transition within the ordered phase of the 3D Ising model.  
This transition appears to be associated with boundary percolation. This is the percolation of dual-plaquettes that lie on the domain boundary between + and - spins, a type of percolation that has not been much studied.  Percolation of domain boundaries occurs when minority sites occupy 13\% or more of the lattice.   It is, incidentally, not coincident with the roughening transition which occurs much deeper into the ordered phase, around $\kappa = 0.408$\cite{rougheningising}.   Because the Ising model has a formulation in terms of the domain boundary itself, the percolation of the boundary could be important, possibly producing a sudden change in the entropy function expressed in terms of boundary area.  

Random boundary percolation appears to have the same critical exponents as ordinary site percolation. Boundary percolation in the Ising model seems to model random boundary percolation as far as the scaling of cluster sizes is concerned, however it differs in the finite-size shift exponent, which determines how the percolation threshold depends on lattice size.  Whereas random percolation has a shift exponent agreeing with typical values of the correlation-length exponent from cluster size scaling ($\nu \sim 0.88$), the shift exponent from the 3D Ising model boundary percolation is vastly different,  $\nu \sim 1.3$.  This surprising result means that the system has two different correlation lengths, both diverging at the infinite-lattice percolation threshold.  This also suggests that there is more than just percolation going on here. If percolation is linked to a thermal phase transition, that could explain the odd shift exponent, since the order parameter of the phase transition may have its own correlation length.

To find such an order parameter we examined the dual system, the 3D gauge Ising model.  The dual point of the boundary percolation threshold occurs in the random (confining) phase of the gauge theory.  An order parameter for the confinement-deconfinement transition can be obtained in Coulomb gauge, where as many one and two direction links as possible are made to be positive by gauge transformations.  The third direction links on each lattice layer can be taken to be a spin-like order parameter, which shows spontaneous magnetization in the ordered phase and is unmagnetized in the random phase.  If there is a phase transition corresponding to boundary percolation in the Ising model itself, it must occur within the random phase of the gauge theory. This suggests looking for a spin-glass transition here, a shift from a completely disordered phase to one with a hidden pattern of order, but still showing no net magnetization.  To this end we utilized a two-real-replica order parameter, which indeed does show a phase transition near the dual reflection of boundary percolation, and with the same critical exponent $\sim 1.30$.  This is significant because it is a true symmetry-breaking phase transition. The symmetry being broken is the layered remnant $(Z2)^L$ symmetry left over after Coulomb gauge fixing, which is global in two dimensions but still local in the third.  This is ``global enough'' to avoid Elitzur's theorem and has sufficient dimensions (2) for a discrete symmetry to break spontaneously at a finite coupling. Spontaneous symmetry-breaking always results in  a phase transition, i.e. a mathematical singularity in the order parameter, which also results in an energy singularity except in a few unusual cases \cite{previous-app2}.  

Finally we examined energy moments in search of this singularity. Because of the high value of $\nu$ the specific heat exponent is negative, implying a finite singularity, so the usual finite-size scaling applied to peak heights cannot be used here.
We concentrated on the third energy cumulant, since it could be fit without the addition of an obfuscating non-singular part, other than a constant.  An open fit to the singular form expected for this quantity based on the hyperscaling relationship, gives $\kappa _c$ and $\nu$ values consistent with those determined by boundary percolation and the dual order parameter.  There is also a noticeable jump in coefficient here, another expectation of this sort of singularity.  Our study did not have enough statistics to see the small expected peak scaling in the fourth cumulant or somewhat larger effect in the fifth, which should have infinite singularities on the infinite lattice.   Although observing these would be satisfying, still the singular fit to the third cumulant does match well with the prediction from the order parameter.  This demonstrates that phase transitions as weak as these can be studied by numerical methods.  The existence of an order parameter and associated symmetry breaking is key in establishing this as a true phase transition.  The coincidence of boundary percolation is also interesting and gives another measure of $\nu$, but cannot by itself be used to imply the presence of a phase transition.  However, it has the advantage of being very easy to measure.  It appears to have the same cluster-size scaling exponents as ordinary site percolation, but a different threshold.  It may be interesting to explore boundary percolation in other systems. Since percolation has so many practical applications, it's possible boundary percolation is a better fit than site or link percolation in some cases. Finally, we note that a previous study of the Z2 gauge-Higgs system showed a total of four phase transition lines further into the diagram\cite{previous}.  The current paper shows there are two phase transitions on each axis, gauge and Higgs, so also a total of four.  It will be interesting to follow these new phase transitions into the phase diagram to see how they connect with the lines previously found.  The previous paper showed that the $Z2$ gauge-Higgs system appears to be more complicated than previously thought. The present paper shows that these additional complications extend to the 3D Ising model itself, and its dual, the 3D gauge Ising model.  It seems possible that similar weak phase transitions may also be lurking in other well-known spin and gauge systems.

\end{document}